\documentclass[aps,prc,reprint,showpacs,groupedaddress,onecolumn]{revtex4-1}

\usepackage{graphicx,color} 
\usepackage{dcolumn}  
\usepackage{bm}       
\usepackage{amsmath} 
\usepackage[dvipsnames]{xcolor}

\begin{document}

\newcommand {\nc} {\newcommand}

\newcommand{\vv}[1]{{$\bf {#1}$}}
\newcommand{\ul}[1]{\underline{#1}}
\newcommand{\vvm}[1]{{\bf {#1}}}
\def\btau{\mbox{\boldmath$\tau$}}

\nc {\IR} [1]{\textcolor{red}{#1}}
\nc {\IB} [1]{\textcolor{blue}{#1}}
\nc {\IP} [1]{\textcolor{magenta}{#1}}
\nc {\IM} [1]{\textcolor{Bittersweet}{#1}}
\nc {\IE} [1]{\textcolor{Plum}{#1}}

\nc{\ninej}[9]{\left\{\begin{array}{ccc} #1 & #2 & #3 \\ #4 & #5 & #6 \\ #7 & #8 & #9 \\ \end{array}\right\}}
\nc{\sixj}[6]{\left\{\begin{array}{ccc} #1 & #2 & #3 \\ #4 & #5 & #6 \\ \end{array}\right\}}
\nc{\threej}[6]{ \left( \begin{array}{ccc} #1 & #2 & #3 \\ #4 & #5 & #6 \\ \end{array} \right) }
\nc{\half}{\frac{1}{2}}
\nc{\numberthis}{\addtocounter{equation}{1}\tag{\theequation}}
\nc{\lla}{\left\langle}
\nc{\rra}{\right\rangle}
\nc{\lrme}{\left|\left|}
\nc{\rrme}{\right|\right|}

\title{\textit{Ab initio}  Folding Potentials for Nucleon-Nucleus Scattering
based on NCSM One-Body Densities} 

\author{M. Burrows$^{(a)}$}
\author{Ch. Elster$^{(a)}$}
\author{S.P. Weppner$^{(b)}$}
\author{K.D. Launey$^{(c)}$}
\author{P.~Maris$^{(d)}$}
\author{A. Nogga$^{(e)}$}
\author{G.~Popa$^{(a)}$}

\affiliation{(a)Institute of Nuclear and Particle Physics,  and
Department of Physics and Astronomy,  Ohio University, Athens, OH 45701,
USA  \\
(b)  Natural Sciences, Eckerd College, St. Petersburg, FL 33711,
USA \\
(c) Department of Physics and Astronomy, Louisiana State University,
Baton Rouge, LA 70803, USA\\
(d) Department of Physics and Astronomy, Iowa State University, Ames, IA 50011, USA \\
(e) IAS-4, IKP-3, JHCP, and JARA-HPC,  Forschungszentrum J\"ulich, D-52428
J\"ulich, GER 
}

\date{\today}

\begin{abstract}
\begin{description}
\item[Background] Calculating  microscopic optical potentials for
elastic nucleon-nucleus scattering has already led to large body of work in the
past. For folding first-order calculations the nucleon-nucleon (NN) interaction and the
one-body density of the nucleus were taken as input to rigorous calculations in
a spectator expansion of the multiple scattering series. 

\item[Purpose] Based on the Watson expansion of the multiple scattering series we
employ a nonlocal translationally invariant nuclear density derived from a
chiral next-to-next-to-leading order (NNLO) 
and the very same interaction for consistent 
full-folding calculation of the effective (optical) potential for nucleon-nucleus
scattering for light nuclei.

\item[Results] We calculate scattering observables, such as total, reaction, and differential cross sections as well as the analyzing power
and the spin-rotation parameter, for elastic scattering of protons and neutrons from
$^4$He, $^{6}$He, $^{12}$C, and $^{16}$O,  in the energy regime between 100 and 200~MeV
projectile kinetic energy, and compare to available data.

\item[Conclusions] Our calculations show that the effective nucleon-nucleus potential obtained from the
first-order term in the spectator expansion of the multiple scattering expansion describes experiments very
well to about 60 degrees in the center-of-mass frame, which coincides roughly with the validity of the NNLO chiral
interaction used to calculate both the NN amplitudes and the one-body nuclear density.

\end{description}
\end{abstract}


\maketitle

\section{Introduction and Motivation}
\label{intro}

Traditionally differential cross sections and spin observables played an important role in either
determining the parameters in phenomenological optical models for proton or neutron
scattering from nuclei or in
testing accuracy and validity of microscopic models thereof. Specifically, elastic scattering of
protons and neutrons from stable nuclei has led in the 1990s to a large body of work on microscopic optical
potentials in which the nucleon-nucleon interaction and the density of the nucleus were taken as
input to rigorous calculations of first-order potentials, in either a Kerman-McManus-Thaler (KMT) or
a Watson expansion of the multiple scattering series 
(see
e.g.~\cite{Crespo:1992zz,Crespo:1990zzb,Elster:1996xh,Elster:1989en,Arellano:1990xu,Arellano:1990zz}),
for which a primary goal was a deeper understanding of the reaction mechanism. 
However, a main disadvantage of that work was the lack of sophisticated nuclear structure input
compared to what is available today. 

Recent developments of the nucleon-nucleon (NN) and three-nucleon (3N) interactions, derived
from chiral effective field theory, have yielded major
progress~\cite{EntemM03,Epelbaum06,Epelbaum:2008ga,Epelbaum:2014sza,Epelbaum:2014efa,Reinert:2017usi,Machleidt:2011zz,Entem:2017gor}. These, together with the
utilization of massively parallel computing resources (e.g.,
see~\cite{LangrSTDD12,SHAO20181,CPE:CPE3129,Jung:2013:EFO}), have placed {\it ab initio}
large-scale simulations at the frontier of nuclear structure and reaction explorations. Among
other successful many-body theories, the {\it ab initio} no-core shell-model (NCSM) approach,
which has considerably advanced our understanding and capability of achieving first-principles
descriptions of low-lying states in light nuclear
systems~(e.g., see~\cite{Navratil:2000ww,Roth:2007sv,BarrettNV13,Binder:2018pgl,BaroniNQ13}), has over
the last decade taken center stage
in the development of microscopic tools for studying the structure of atomic nuclei. The NCSM
concept combined with a symmetry-adapted (SA) basis in the {\it ab initio} SA-NCSM
\cite{DytrychLMCDVL_PRL12} has further expanded the reach to the structure of intermediate-mass
nuclei \cite{LauneyDD16}. 

\begin{figure}
\centering
\includegraphics[width=10cm]{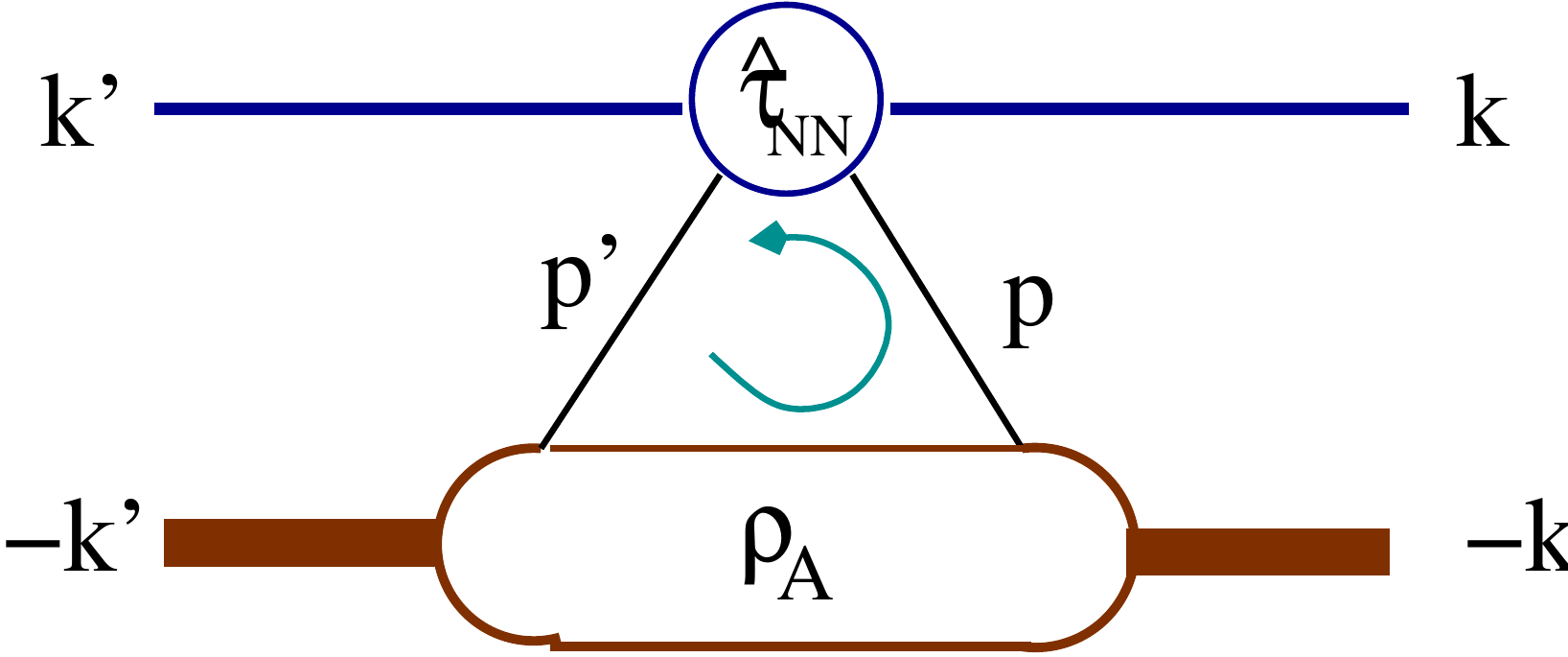}
\caption{Diagram for the matrix element of the effective (optical) potential for
the single scattering term.
}
\label{fig1}
\end{figure}

Following these developments in nuclear structure theory, it is worthwhile to
again consider rigorous calculations of effective folding nucleon-nucleus (NA) potentials,
since now the one-body densities required for the folding with NN scattering amplitudes can be
based on the same NN interaction, and thus can be considered {\it ab initio}. This is complementary  to recent developments, where effective NA potentials are extracted from {\it ab initio} structure calculations via Green's function methods~\cite{Rotureau:2016jpf}. Our approach to elastic scattering is based on the spectator expansion of multiple scattering theory~\cite{Ernst:1977gb,Siciliano:1977zz,Tandy:1980zz,Chinn:1995qn}. 
Here the first-order term involves two-body interactions between the projectile and one of the
target nucleons which requires a convolution of the fully off-shell NN scattering amplitude with the nuclear wave functions of
the target represented by a nonlocal one-body density (OBD). Thus, in its most general form,
the first-order single scattering optical potential within the framework of the spectator
expansion is given by the triangle graph shown in Fig.~\ref{fig1}. A specific scope of this
work is to consistently obtain the NN scattering amplitudes and the nuclear one-body densities
from a chiral NN interaction up to next-to-next-leading order. We neglect the three-nucleon forces (3NFs) in this work since they are known to only give small contributions to densities and do not contribute to the Watson expansion in the first order of the optical interaction.
Similar  work  in  this direction  is  carried  out  in  Ref.~\cite{Gennari:2017yez},  however using a different chiral NN
interaction~\cite{Entem:2017gor} for the NN scattering amplitudes, which is augmented by 3N
interaction and is renormalized in calculations of the nuclear density. It is interesting to compare the results of this work to those in Ref.~\cite{Gennari:2017yez}.

The structure of the paper is as follows: In Sec.~\ref{formal} we review the formalism for the
single-scattering folding potential and introduce the full-folding procedure as used in our
calculations. Though in principle this can be found in the literature, for clarity and the convenience
of the reader we give the most important steps here.
 In Sec.~\ref{results} we present results for elastic scattering of protons as
well as neutrons from the ``closed shell" nuclei $^4$He and $^{16}$O in the energy regime between
100 and 200~MeV. Then we apply the
formulation to the ``open shell" nuclei $^{12}$C and $^6$He. Our conclusions are presented in
Sec.~\ref{conclusions}.


\section{The First-order Folding Potential}
\label{formal}

The standard approach to elastic scattering of a strongly interacting
projectile from
a target of $A$ particles is the separation of the Lippmann-Schwinger (LS)
equation for the transition amplitude
\begin{equation}
T = V + V G_0(E) T  \label{eq:2.1}
\end{equation}
into two parts, namely an integral equation for $T$:
\begin{equation}
T = U + U G_0(E) P T  , \label{eq:2.2}
\end{equation}
where $U$ is the effective (optical) potential operator and defined by a second
integral equation
\begin{equation}
U = V + V G_0(E) Q U.  \label{eq:2.3}
\end{equation}
In the above equations the operator $V$ represents the external
interactions between the projectile and the target nucleons, and the projection  operators $P$ and $Q$ are defined below. 
The Hamiltonian for the
$(A+1)$-particle system is given by
\begin{equation}
H=H_{0}+V .  \label{eq:2.4}
\end{equation}
The potential operator $V=\sum_{i=1}^A v_{0i}$ consists of the NN potential $v_{0i}$ acting
between the projectile, denoted by ``0", and the $i$-th target nucleon. The free propagator for the projectile+target
system is given by $G_0(E)=(E-H_0+i\varepsilon)^{-1}$, where $H_0=h_0+H_A$, with
$h_0$ being the kinetic energy operator for the projectile and $H_A$ denoting the target Hamiltonian.
Defining $|\Phi_{A}\rangle$ as the ground state of the target, we have $H_A |\Phi_{A}\rangle =E_A
|\Phi_{A}\rangle$. The operators $P$ and $Q$ in Eqs.~(\ref{eq:2.2}) and (\ref{eq:2.3}) are projection
operators, $P+Q=1$, and $P$ is defined such that Eq.~(\ref{eq:2.2}) is
solvable. In this case, $P$ is conventionally taken to project onto
the elastic channel, such that $[G_{0},P]=0$, and is given as
$P=\frac{|\Phi_{A}\rangle\langle\Phi_{A}|}{\langle\Phi_{A}|\Phi_{A} \rangle}$. 
With these definitions
the transition operator for elastic scattering
can be defined as ${T_{el}=PTP}$, in which case Eq.~(\ref{eq:2.2})
 becomes
\begin{equation}
T_{el}=PUP + PUPG_{0}(E)T_{el}.  \label{eq:2.5}
\end{equation}

The fundamental idea of the spectator expansion for the optical
potential is an ordering  of the scattering process according to the
number of active target nucleons interacting directly with the
projectile. The first-order term involves two-body interactions between
the projectile and one of the target nucleons, {\it i.e.} $U =
\sum_{i=1}^{A}\tau_{i}$, where the operator $\tau_{i}$ is derived to be
\begin{eqnarray}
\tau_i &=& v_{0i} + v_{0i} G_0(E) Q \tau_i \nonumber \\
&=& v_{0i} + v_{0i}G_0(E) \tau_i - v_{0i} G_0(E) P \tau_i
\label{eq:2.6} \\
&=& \hat{\tau_i} - \hat{\tau_i} G_0(E) P \tau_i . \nonumber
\end{eqnarray}
Here $\hat{\tau_i}$ is the NN t-matrix and is defined as the solution of
\begin{equation}
\hat{\tau_i} = v_{0i} + v_{0i} G_0(E) \hat{\tau_i}. \label{eq:2.7}
\end{equation}
It should be noted that all of the above equations follow in a straightforward derivation and correspond
to the first-order Watson scattering expansion \cite{Watson1953a,Watson1953b}. In the lowest order 
the operator $\hat{\tau_i}\approx t_{0i}$, which corresponds to the conventional impulse
approximation. Here the operator $t_{0i}$ stands for the standard solution of a 
LS equation with the NN interaction as driving term. It should be pointed out that the implicit treatment
of the operator $Q$ in Eq.~(\ref{eq:2.6}) is especially important for scattering from light nuclei as
shown in Ref.~\cite{Chinn:1993zza}.

For elastic scattering only $P\tau_i P$ (from Eq.~(\ref{eq:2.6})) needs to be considered, or equivalently
\begin{equation}
\langle\Phi_A| \tau_i | \Phi_A\rangle = \langle\Phi_A| \hat{\tau_i}|
 \Phi_A\rangle - \langle\Phi_A| \hat{\tau_i}| \Phi_A\rangle \frac {1}
 {(E-E_A) - h_0 + i\varepsilon} \langle\Phi_A| \tau_i | \Phi_A\rangle ,
 \label{eq:2.8}
\end{equation}
and this matrix element represents the full-folding effective (optical)
potential
\begin{equation}
\langle{\bf k}' | U |{\bf k}\rangle =
\langle{\bf k}'\Phi_{A} | \sum_{i} {\tau_{i}}
|{\bf k}\Phi_{A}\rangle , \label{eq:2.9}
\end{equation}
Since $\langle{\bf k}' | U |{\bf k}\rangle$ is the solution of the sum
of one-body integral equations represented by Eq.~(\ref{eq:2.8}),
it is sufficient to consider the driving term
\begin{equation}
\langle{\bf k}' |\hat{U}|{\bf k}\rangle =
\langle{\bf k}'\Phi_{A} | \sum_{i} \hat{\tau}_{i}
|{\bf k}\Phi_{A}\rangle , \label{eq:2.10}
\end{equation}
where $\hat{\tau}_{i} \approx t_{0i}$  in the impulse approximation. 
Inserting a complete set of momenta for the struck target nucleon
before and after the collision and representing the sum over target protons and neutrons by $\alpha$ leads to
\begin{eqnarray}
\hat{U}\left({\bf k^{\prime}},{\bf k}\right)=
\sum_{\alpha=p,n} \int d^{3}{\bf p^{\prime}}
d^{3}{\bf p} \left\langle {{\bf k^{\prime}}{\bf p^{\prime}}}
\mid \hat{\tau}_{\alpha} (\epsilon)
\mid {{\bf k p}} \right\rangle \rho_{\alpha} \left({{\bf p^{\prime}}
+\frac{{\bf k'}}{A}}, {\bf p}+\frac{{\bf k}}{A}\right)
\delta^{3} ({\bf k^{\prime}} + {\bf p^{\prime}} -
{\bf k} - {\bf p}), \label{eq:2.11}
\end{eqnarray}
where the momenta ${\bf k'}$ and ${\bf k}$ are the final and initial momenta
of the projectile in the frame of zero total nucleon-nucleus
momentum. The structure of Eq.~({\ref{eq:2.11}}) is represented
graphically by Fig.~\ref{fig1}, which also illustrates the momenta ${\bf p'}$ and
${\bf p}$. The proton and neutron densities are given
by $\rho_{\alpha}$. Evaluating the $\delta$-function, introducing the variables 
${\bf q}={\bf k'}-{\bf k}$, ${\bf K}=\frac{1}{2}({{\bf k} + {\bf k^{\prime}}})$ and
${\bf \hat{p}}=\frac{1}{2} ({{{\bf p^{\prime}}}+{{\bf p}}})$, and finally changing the integration
variable from ${\bf \hat{p}}$ to ${\bf P}={\bf\hat p}
+ \frac{{\bf K}}{A}$, accounting for the recoil of the nucleus~\cite{Picklesimer:1984bb},
leads to the final expression for the full-folding effective potential
\begin{eqnarray}
\hat{U}({\bf q},{\bf K})=\sum_{\alpha=p,n}
\int d^{3}{\bf P}\;
\eta({\bf P},{\bf q},{\bf K})\;
\hat{\tau}_{\alpha}\left({\bf q},\frac{1}{2}\left(\frac{A+1}{A}{\bf K}-{\bf P}\right);
\epsilon\right)\; \nonumber \\
\;\;\;\;\;\;\;\;\;\;\;\;\;\;\;
\times \; \rho_{\alpha}\left({\bf P}-\frac{A-1}{A}\frac{\bf q}{2},
{\bf P}+\frac{A-1}{A}\frac{\bf q}{2}\right). \label{eq:2.13}
\end{eqnarray}
Here $\eta({\bf P},{\bf q},{\bf K})$ is the M\o ller factor for the frame
transformation~\cite{Joachain} relating the NN zero-momentum frame to the NA zero-momentum frame.
Further details can be found in Refs.~\cite{Picklesimer:1984bb,Elster:1996xh,Weppner:1997}.
The free NN amplitude ${\hat \tau}_\alpha$ is calculated from the free NN t-matrix according to 
Eq.~(\ref{eq:2.7}) at an appropriate energy $\epsilon$. 
In principle this energy should be the beam energy minus the kinetic energy of the center-of-mass
(c.m.)
of the interacting particle less the binding energy of the struck particle. Following this argument,
$\epsilon$ should be coupled to the integration variable ${\bf P}$. The full-folding calculations 
of Refs.~\cite{Arellano:1994zz,Elster:1997as} are carried out along this vein, and found only small
effects for scattering energies above 100~MeV. For our calculation we take a different approach, we
fix $\epsilon$ at the two-body c.m. energy corresponding to the free NN scattering at the beam energy
so that the same laboratory energy applies to the two-body and nuclear scattering. This approach has
also been applied in earlier work~\cite{Elster:1996xh,Chinn:1994xz,Chinn:1993zza,Elster:1989en}.
The quantity  $\rho_{\alpha}$, with $\alpha = p(n)$, represents a nonlocal OBD for the proton (neutron)
distribution. Since $\hat{U}({\bf q},{\bf K})$ is computed in the NA c.m. frame, it is mandatory that the
OBD must be given in a translationally invariant fashion.

An important product of this work is that the NN t-matrix and OBD now use the same underlying
NN interaction.
For this we choose the optimized chiral NN interaction at the
next-to-next-to-leading order NNLO$_{\rm{opt}}$ from Ref.~\cite{Ekstrom13}. This interaction is
fitted with $\chi^2 \approx 1$ per degree of freedom for laboratory energies up to about 125~MeV. In the
$A$~=~3, 4 nucleon systems the contributions of the 3NFs are smaller than in most other parameterizations of
chiral interactions. As a consequence, nuclear quantities like root-mean-square radii
and electromagnetic transitions in light and intermediate-mass nuclei can be calculated  reasonably well without invoking 3NFs \cite{Henderson:2017dqc,LauneyMSSBMDD18}.
 From this point of view, the NNLO$_{\rm{opt}}$ 
 NN interaction is very well suited for
our calculations, since the first-order folding potential does not contain any explicit 3NF contributions.

The full-folding effective potential of Eq.~(\ref{eq:2.13}) requires as input a nonlocal translationally
invariant OBD. The procedure for computing this quantity from {\it ab initio} NCSM calculations has
been described in detail in Ref.~\cite{Burrows:2017wqn}, and the derivation will not be repeated here. 
The convolution of the nonlocal OBD with the fully off-shell NN t-matrix and the M\o ller frame
transformation factor is carried out in momentum space 
in three dimensions without partial wave decomposition, and the
integration is performed using Monte Carlo integration techniques. It is also to be understood that all
spin summations are performed in obtaining ${\hat {U}}({\bf q},{\bf K})$.  For a strictly
spin-zero nucleus, this reduces the required NN
t-matrix elements to a spin-independent component (corresponding to the Wolfenstein amplitude $A$) and a
spin-orbit component (corresponding to Wolfenstein amplitude $C$), whereas the components of the NN t-matrix depending on the spin of the
struck nucleon vanish. For the proton nucleus scattering calculations the Coulomb interaction between the projectile
and the target is included using the exact formulation from Ref.~\cite{Chinn:1991jb}.

\begin{figure}
\centering
\includegraphics[width=13cm]{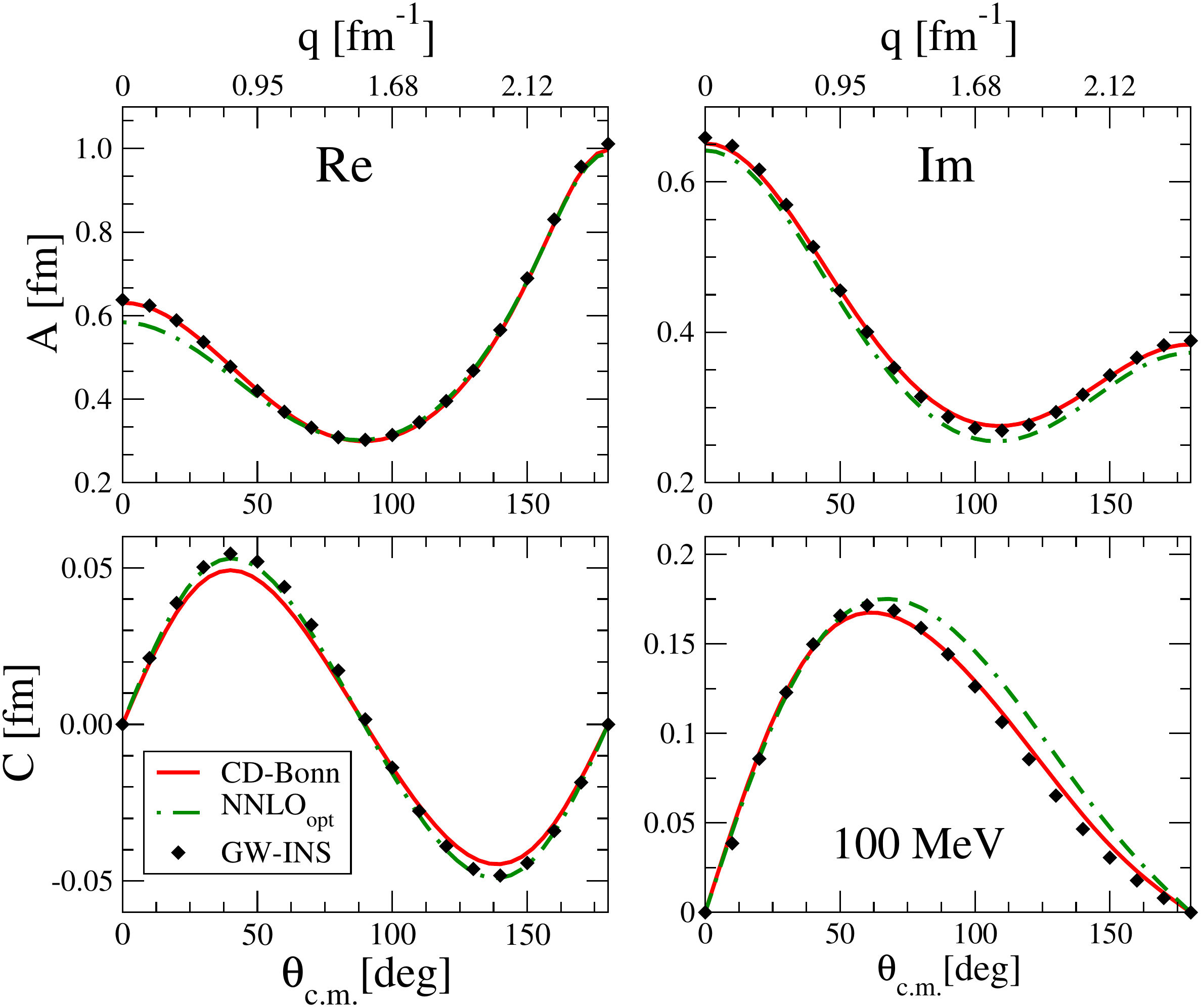}
\caption{Wolfenstein amplitudes A and C as function of the scatting angle and
momentum transfer for $np$ scattering at 100~MeV laboratory kinetic energy. The
solid (red) line stands for the CD-Bonn potential~\protect\cite{Machleidt:2000ge} and the
dash-dotted (green) line for the NNLO${_{\rm{opt}}}$ chiral
interaction~\protect\cite{Ekstrom13}. The solid diamonds represent the extraction from the
GW-INS analysis~\protect\cite{Workman:2016ysf}.
}
\label{fig2}
\end{figure}

\begin{figure}
\centering
\includegraphics[width=13cm]{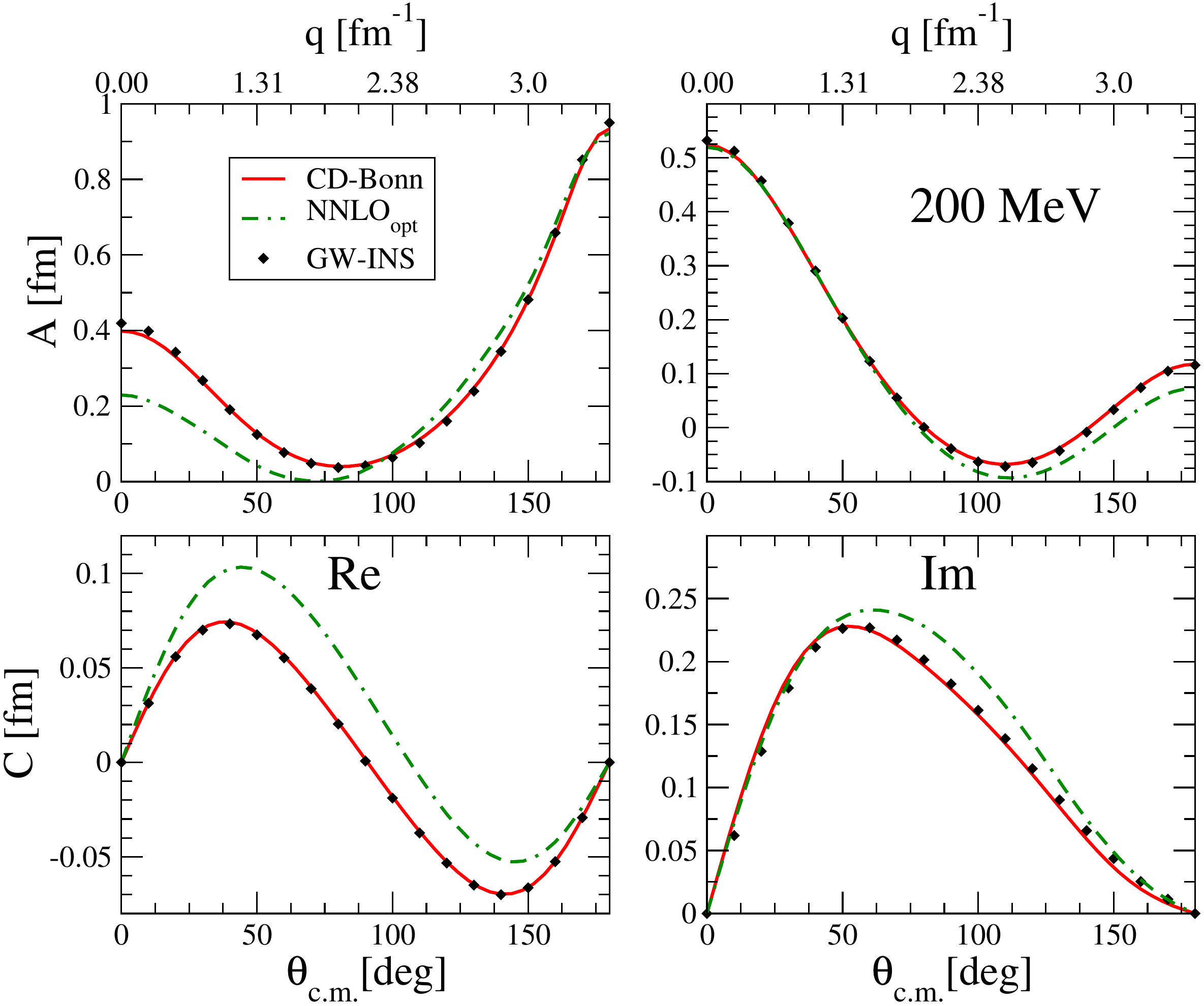}
\caption{Same as Fig.~\ref{fig3} for $np$ scattering at 200~MeV laboratory kinetic
energy.
}
\label{fig3}
\end{figure}

Since our calculations for NA scattering concentrate on the energy regime between ~100 and 200~MeV, we
first want to consider how well the Wolfenstein amplitudes $A$ and $C$ are described by the chiral NN
interaction NNLO$_{\rm{opt}}$. This comparison is shown in Fig.~\ref{fig2} for 100~MeV and
Fig.~\ref{fig3} for 200~MeV for the $np$ Wolfenstein amplitudes. 
All figures show $A$ and $C$ obtained from NNLO$_{\rm{opt}}$ together with
the experimental extraction from the GW-INS analysis~\cite{Workman:2016ysf}. As comparison we also show
$A$ and $C$ obtained from the Charge-Dependent Bonn potential (CD-Bonn)~\cite{Machleidt:2000ge}, 
which is fitted to the NN data up to 300~MeV with $\chi^2\approx 1$. As expected at 100~MeV NN laboratory
kinetic energy differences between NNLO$_{\rm{opt}}$, CD-Bonn, and the experimental extraction from the
GW-INS analysis are minimal. The imaginary part of Wolfenstein $C$ determines the real part of the NA
spin-orbit interaction. 
The NNLO$_{\rm{opt}}$ interaction will result in a slightly
stronger spin-obit term (related to $\Im m \;C$) above 100~MeV. Likewise the real part of A 
(the central depth) in the forward
direction is slightly under-predicted by NNLO$_{\rm{opt}}$ at 100 MeV and becomes strongly under-predicted by 200 MeV. 
The differences exhibited by  NNLO$_{\rm{opt}}$ changes the ratio between the central depth, and the spin-orbit force, an
important factor in the spin observables in NA scattering. 
This disparity may be a consequence of the interaction having a small $\chi^2$ below 125~MeV NN laboratory kinetic
energy, while by 200~MeV the $\chi^2$ is about 6 in the $np$ channel, 
with the largest disagreement being in the $P$-waves.


\section{Results and Discussion}
\label{results}

\subsection{Elastic scattering observables for $^4$He and $^{16}$O}

The first-order folding potential for NA scattering, as described in the previous
section, is exact for nuclear states with total intrinsic spin zero, so
we first concentrate on ``closed shell" nuclei, such as $^4$He and
$^{16}$O, with a ground state that is largely dominated by zero intrinsic spin.
For example, converged cross section results for $^4$He, discussed below, use NCSM calculations of the $^4$He ground state that has a spin-zero contribution of about 95\%. 
The ``closed shell" nuclei within the reach of NCSM calculations are $^4$He and
$^{16}$O. After computing the first-order folding potential using as input a nonlocal translationally
invariant OBD based on the NNLO$_{\rm{opt}}$ chiral potential~\cite{Ekstrom13}
obtained as outlined in Ref.~\cite{Burrows:2017wqn} and the Wolfenstein amplitudes $A$ and $C$ based on the
same interaction, we compute total, reaction, and differential cross sections for elastic scattering as well as the analyzing power $A_y$
and the spin-rotation parameter $Q$. Our choice of energies for which we show observables is dictated by the
availability of experimental data, and we concentrate on the energy regime between 100 and 200~MeV projectile
laboratory kinetic energy since we expect that the first-order term governs the scattering process at those
energies.

\begin{figure}
\centering
\includegraphics[width=17cm]{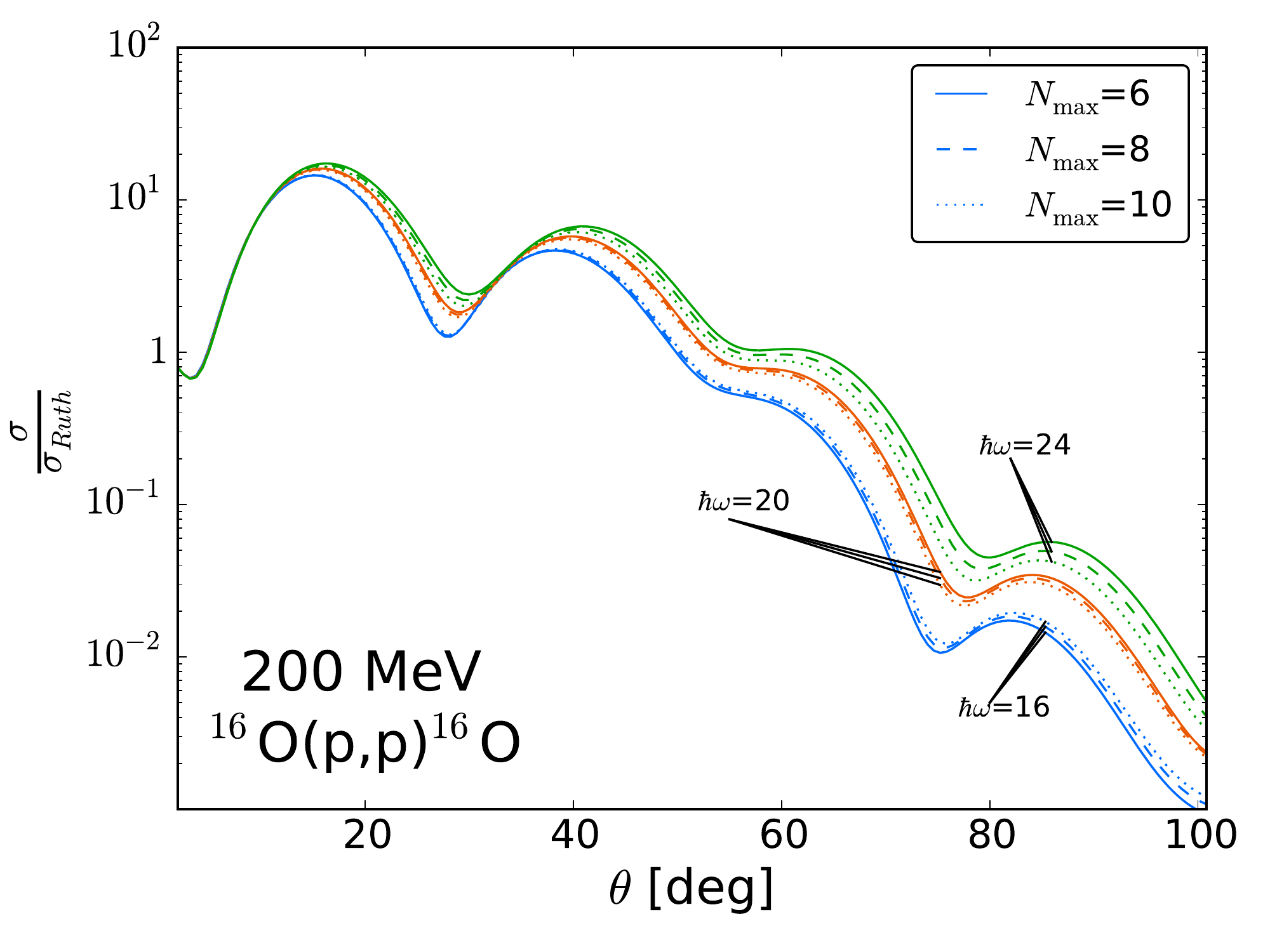}
\caption{The angular distribution of the differential cross section divided by the
Rutherford cross section for elastic proton scattering from $^{16}$O at 200~MeV 
laboratory kinetic energy as function of the c.m. angle calculated with the NNLO${_{\rm{opt}}}$ chiral
interaction~\protect\cite{Ekstrom13}. The different values of $N_{\rm{max}}$ are indicated in the legend. 
From top to bottom, the three sets of lines correspond to $\hbar\omega$~=~24, 20, and 16~MeV respectively.
}
\label{O16_Nmax}
\end{figure}

The nonlocal translationally invariant densities are calculated from one-body density matrix elements computed in the NCSM framework. The latter uses a harmonic-oscillator basis characterized by two parameters, 
$N_{\max}$, defined as the maximum number of oscillator quanta above the valence shell for that nucleus as
well as the oscillator length $\hbar\omega$. A converging trend of nuclear structure observables, including binding energies and radii, with respect to these model parameters has been ensured but this does not necessarily ensure convergence of the scattering observables under consideration, details of which we present herein.
It is well known that different observables exhibit a different convergence behavior with respect to the two parameters. While the scattering observables presented here for $^4$He are well converged already at $N_{\max}=8$ and practically independent of  $\hbar\omega$ over the range of 16-24 MeV (further discussed below for $N_{\max}=18$), in Fig.~\ref{O16_Nmax} we show results for $^{16}$O as an illustrative example,  and we investigate the convergence of the ratio of the differential cross section to the Rutherford cross section at 200~MeV with respect to $N_{\max}$ for three values of $\hbar\omega=16,$~20, and 24~MeV. Here $N_{\max}$~=~6, 8, and 10 results are shown to indicate that the calculations in the $\hbar\omega$ range of 16-20 MeV are almost converged at $N_{\max}$~=~10, with the results for $\hbar\omega=16$~MeV and $\hbar\omega=20$~MeV slowly approaching each other. The results in Fig.~\ref{O16_Nmax} show that the dependence on a selected $\hbar\omega$ range dominates variations in the calculated observables, which is why in the following calculations of scattering observables we only show results across various $\hbar\omega$ values, while keeping $N_{\max}$ at a fixed, reasonably large value.

\begin{figure}
\centering
\includegraphics[width=10cm]{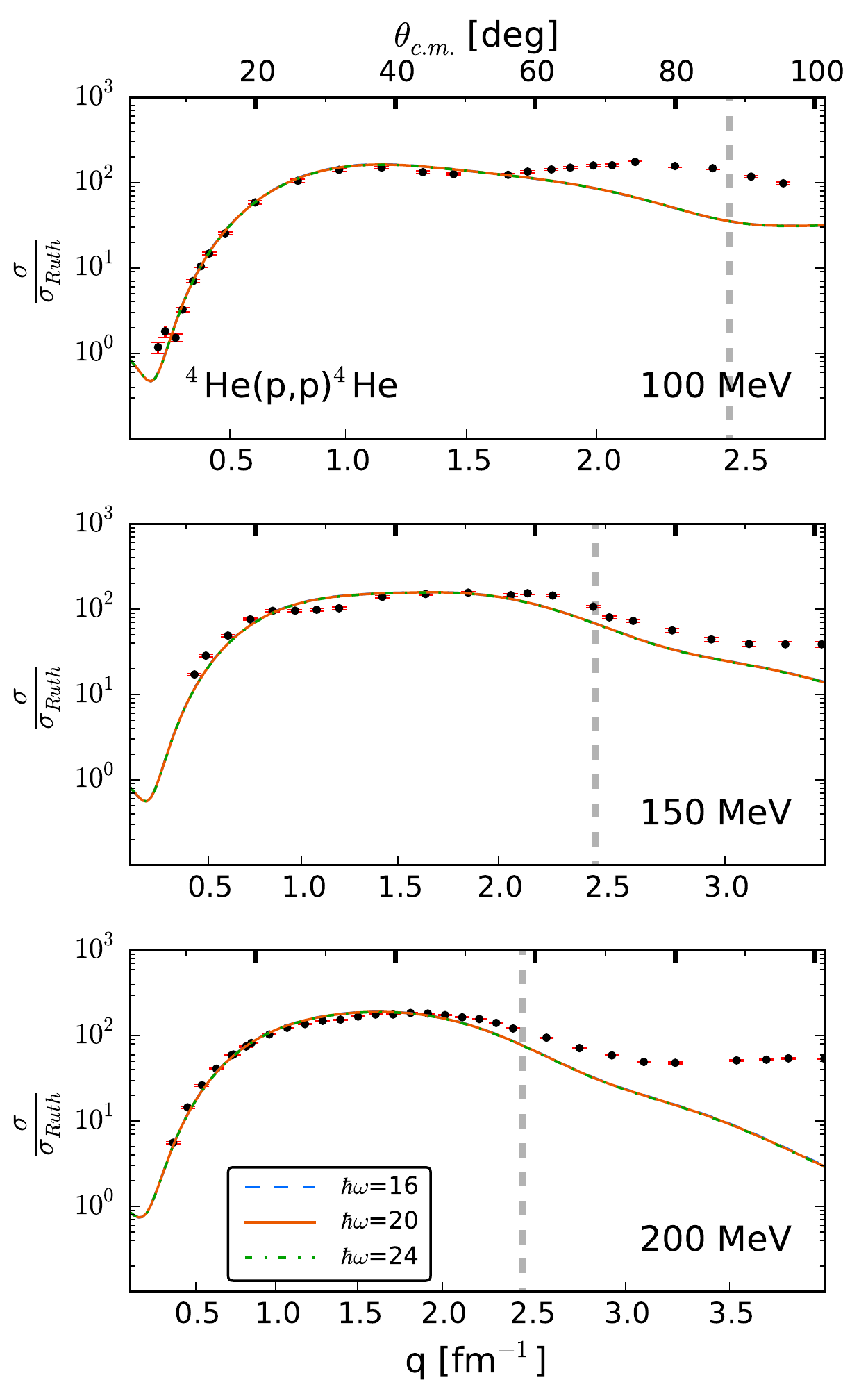}
\caption{The angular distribution of the differential cross section divided by the
Rutherford cross section for elastic proton scattering from $^4$He at 100, 150, and
200~MeV laboratory kinetic energy as function of the momentum transfer and the c.m.
angle calculated with the NNLO${_{\rm{opt}}}$ chiral
interaction~\protect\cite{Ekstrom13}. The dashed line represents the calculation based
on nonlocal densities using
$\hbar\omega$~=~16~MeV, the solid line with 20~MeV, and the dash-dotted line with
24~MeV. For all calculations $N_{\rm{max}}$~=~18 is employed. The data for 100~MeV
are taken from~Ref.~\cite{Goldstein:1970dg}, for 156~MeV from Ref.~\cite{Comparat:1975bm}, and for 200~MeV from Ref.~\cite{Moss:1979aw}. 
The dashed vertical line in each figure indicates the momentum transfer
$q=2.45$~fm$^{-1}$ corresponding to the laboratory kinetic energy 125~MeV of the $np$
system.
}
\label{fig4}
\end{figure}

The differential cross section divided by the Rutherford cross section is shown for scattering of 
protons off $^{4}$He in Fig.~\ref{fig4} for three projectile laboratory kinetic energies, 100, 150, and 
200~MeV as function of the momentum transfer as well as of the c.m. scattering angle. 
Dividing by the Rutherford cross section allows for a clearer view of the forward angles, which
should be well described by the first-order folding potential. This is indeed the case, Fig.~\ref{fig4} shows
that in the energy regime between 100 and 200~MeV the differential cross section is very well described
by the calculations up to about 60$^{\circ}$. At larger angles multiple scattering effects, which are not included,
are likely to become more important. This is a well known phenomenon in, e.g., three-body scattering, where
higher-order Faddeev terms are needed to build up the backward angles in neutron-deuteron
scattering~\cite{Gloeckle:1995jg,Elster:2008yt}.
The vertical dashed line marks the momentum transfer $q=2.45$~fm$^{-1}$ which corresponds to the laboratory
kinetic energy of 125~MeV in the $np$ system, up to which the chiral NNLO$_{\rm{opt}}$ interaction was
fitted.
The cross sections are shown at $N_{\max}=18$ for three different oscillator parameters $\hbar\omega=16$,~20, and 24~MeV, indicating no dependence on the model parameters for this $\hbar\omega$ range. Indeed, for
$N_{\rm max}=18$ the variation in the calculated cross sections with different $\hbar\omega$ values for $^4$He is smaller than the curve widths.

\begin{figure}
\centering
\includegraphics[width=10cm]{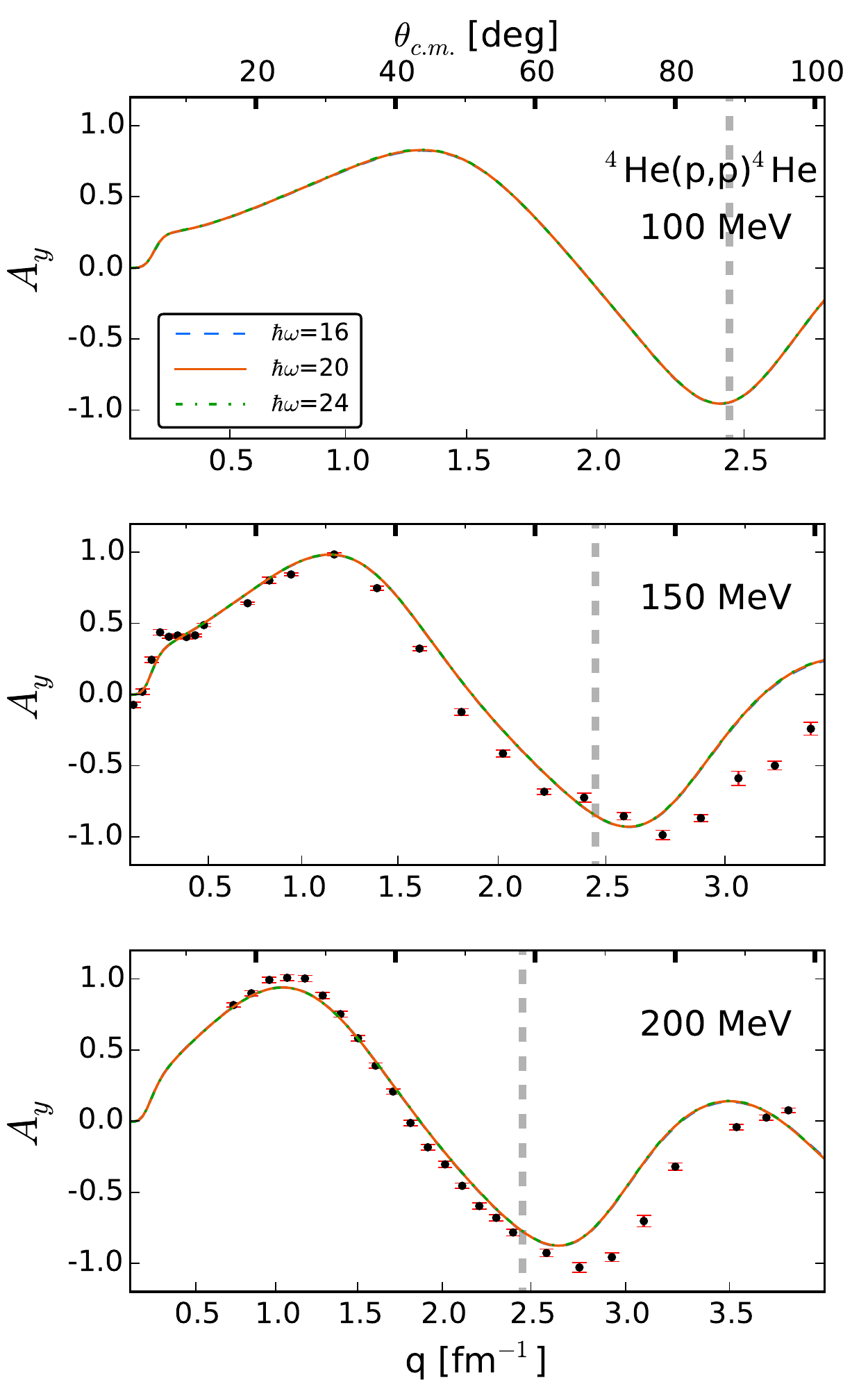}
\caption{The angular distribution of the analyzing power for elastic proton
scattering from $^4$He at 100, 150, and
200~MeV laboratory kinetic energy as function of the momentum transfer and the c.m.
angle calculated with the NNLO${_{\rm{opt}}}$ chiral
interaction~\protect\cite{Ekstrom13}. The lines follow the same notation as in
Fig.~\ref{fig4}. The data for 150~MeV are taken from Ref.~\cite{Cormack:1959zz}, and for 200~MeV from
Ref.~\cite{Moss:1979aw}.
}
\label{fig5}
\end{figure}

The corresponding analyzing power $A_y$ of elastic proton scattering off $^4$He at 100, 150, and 200 MeV laboratory kinetic energy are shown in Fig.~\ref{fig5}. For 150 and 200 MeV, the analyzing power has a reasonably good agreement up to 60$^{\circ}$ and at the line marker. Varying oscillator parameters $\hbar\omega$ at $N_{\max}=18$ produces a very small difference in the calculated cross section, that is smaller than the curve widths shown. This is quite different from the calculations presented in Ref.\cite{Gennari:2017yez}, where the analyzing power of $^{4}$He at 200 MeV misses most data by a considerable amount. In part, our better agreement may be due to our treatment of the projection operator $Q$ as outlined in Eqs.~(\ref{eq:2.6}-\ref{eq:2.8}), which is important for scattering from light nuclei \cite{Rotureau:2016jpf}. Another possibility may be the choice of the underlying NN interaction leading to a very different spin-orbit force. This will need to be further explored.

\begin{figure}
\centering
\includegraphics[width=10cm]{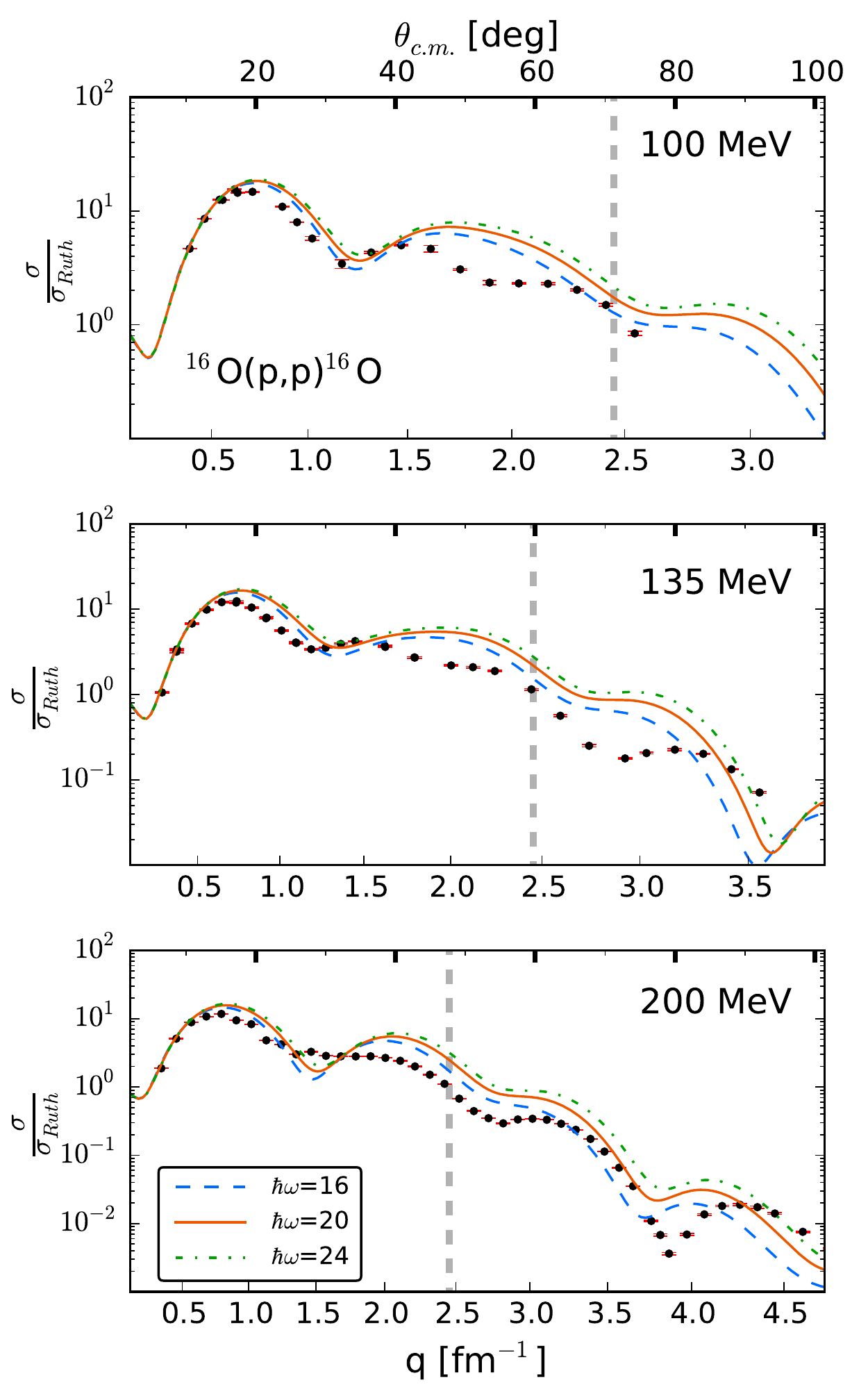}
\caption{The angular distribution of the differential cross section divided by the
Rutherford cross section for elastic proton scattering from $^{16}$O at 100, 135, and
200~MeV laboratory kinetic energy as function of the momentum transfer and the c.m.
angle calculated with the NNLO${_{\rm{opt}}}$ chiral
interaction~\protect\cite{Ekstrom13}. The dashed line represents the calculation based
on nonlocal densities using
$\hbar\omega$~=~16~MeV, the solid line with 20~MeV, and the dash-dotted line with
24~MeV. For all calculations $N_{\rm{max}}$~=~10 is employed. The data for 100~MeV are taken from Ref.~\cite{Seifert:1990um}, for 135~MeV from Ref.~\cite{Kelly:1989zza}, and for 200~MeV from Ref.~\cite{Glover:1985xd}.
The dashed vertical line in each figure indicates the momentum transfer
$q=2.45$~fm$^{-1}$ corresponding to the laboratory kinetic energy 125~MeV of the $np$
system.
}
\label{fig6}
\end{figure}

\begin{figure}
\centering
\includegraphics[width=9.5cm]{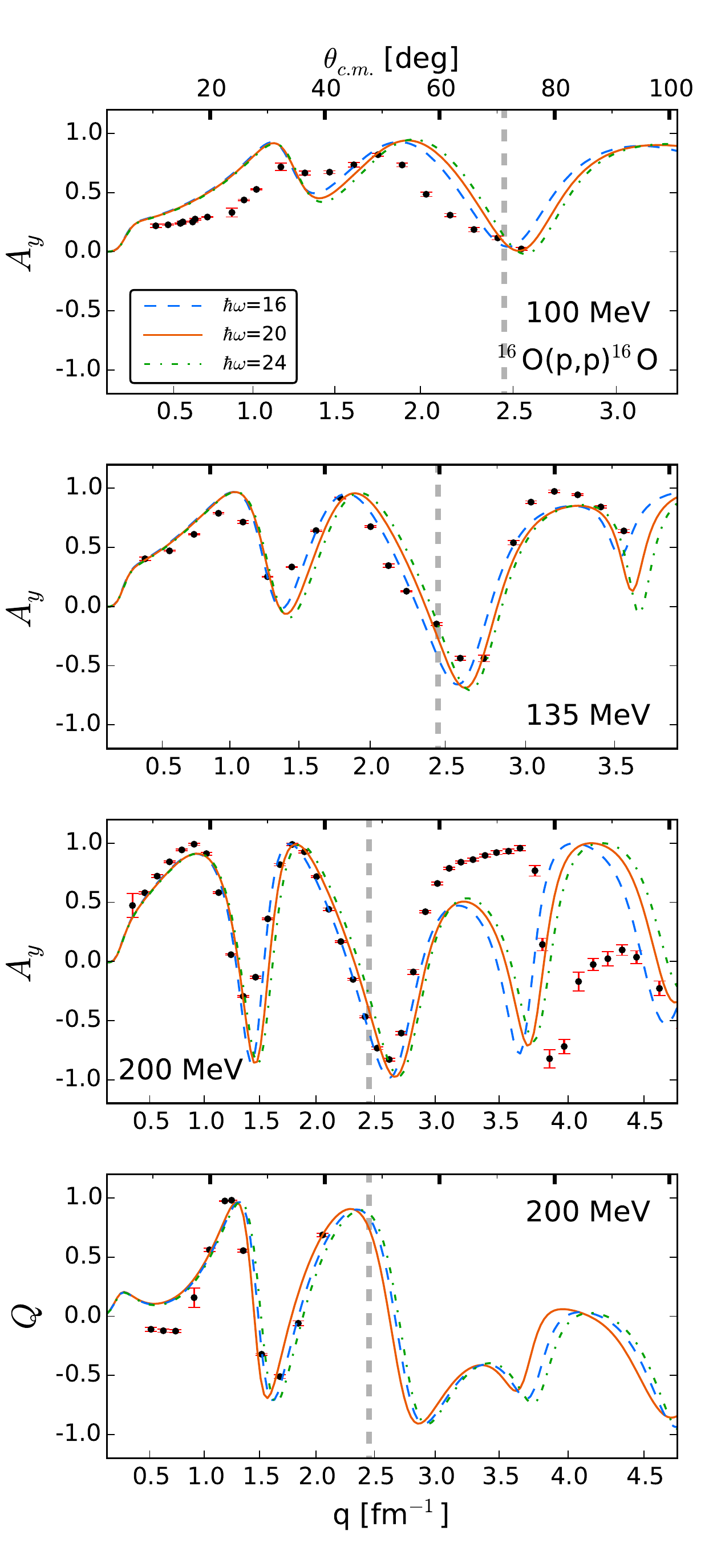}
\caption{The angular distribution of the analyzing power for elastic proton
scattering from $^{16}$O at 100, 135, and
200~MeV laboratory kinetic energy as function of the momentum transfer and the c.m.
angle calculated with the NNLO${_{\rm{opt}}}$ chiral
interaction~\protect\cite{Ekstrom13}. Also included is the angular distribution of the spin rotation parameter for elastic proton scattering from $^{16}$O at 200~MeV. The lines follow the same notation as in
Fig.~\ref{fig6}. The data for 100~MeV are taken from Ref.~\cite{Seifert:1990um}, for 135~MeV from Ref.~\cite{Kelly:1989zza}, and for 200~MeV from
Ref.~\cite{Glover:1985xd}.
}
\label{fig7}
\end{figure}

The calculations of the differential cross section divided by the Rutherford cross section for proton elastic scattering off $^{16}$O is shown in Fig.~\ref{fig6}. The analyzing power for laboratory kinetic energies 100, 135, and 200 MeV are shown in Fig.~\ref{fig7}. Similar to the calculations for $^{4}$He, the value of $N_{\rm max}$ is kept constant, in this case at $N_{\max}=10$, which is the largest $N_{\max}$ achievable in the NCSM with current resources, while $\hbar \omega$ is varied between 16 and 24~MeV. The agreement between the calculated
 differential cross section and the data is reasonable at forward angles (up to 40$^{\circ}$) and low momentum transfer with deviations beginning at around 1.5 to 2 fm$^{-1}$ at all energies. The dependence of the differential cross section on the basis $\hbar\omega$ values indicates that the calculations are not yet fully converged at $N_{\max}=10$. However, at small angles corresponding to low values of the momentum transfer $q$, where we agree reasonably well with the data, this dependence is relatively small.

The experimental data for the analyzing power for $^{16}$O are quite well described for proton energies 135 and 200~MeV for
momentum transfers $q\leq 2.45$~fm$^{-1}$ (Fig.~\ref{fig7}). Here again, the analyzing power shows a weak dependence on $\hbar\omega$ at small angles (low momentum transfer), but this dependence increases with the scattering angle. In fact, $A_y$ is better described than the differential cross
section, indicating that the ratio between central and spin-orbit force is still captured by the calculation
while the absolute magnitude starts to deviate with increasing angles or momentum transfers.
The comparison to experimental data at 100~MeV shows the same general shape but the agreement is not quite the same as the one observed
at higher energies.
This is most likely an indication that higher-order terms in the spectator expansion may become more
important at lower energies. Included in Fig.~\ref{fig7} is also the spin rotation parameter 
 at 200~MeV. Like the analyzing power at the same energy, good agreement between the experimental data 
and the calculation is obtained. A comparison to earlier calculations of the full-folding microscopic
 potential \cite{Chinn:1995qn} shows 
improvement in both the differential cross section and the analyzing power for a larger range of angles. Note that the region below $q=2.45$~fm$^{-1}$ is the region where NNLO$_{\rm opt}$ was fitted, and this is the region where we have reasonably good convergence and agreement with the data. Again, comparing with Ref.\cite{Gennari:2017yez} reveals that our calculations describe the experimental values much better, indicating that the spin-orbit force of Ref.\cite{Gennari:2017yez} is quite different from our calculations.


\subsection{Elastic scattering observables for $^{12}$C and $^{6}$He}

Strictly speaking the full-folding implementation of the first-order term in the multiple scattering
expansion is exact only for nuclear states with a zero intrinsic spin, since -- by definition -- spin-dependent terms in the first-order folding potential that involve a spin flip of the struck target nucleon naturally vanish for a spin-zero state of the target. We note, however, that besides omitting these spin-dependent terms, the present formalism is valid for a general nuclear state with a mixture of any intrinsic spins. To investigate the quality of describing scattering observables using this formalism, we want to consider ``open shell" even-even nuclei. These nuclei have a ground state that is dominated by spin zero and often the spin-zero component is found to be in excess of 80\% of the total wave function (e.g., see Table 3 in Ref. \cite{LauneyDD16} for calculations using NNLO$_{\rm opt}$ and another realistic interaction). 
For example, for $^{6}$He, calculations at $N_{\max}=12$ show that the zero-spin contribution to the ground state is about 80-85\%.
An interesting case is $^{12}$C, for which the ground state has a comparatively large non-zero spin component, namely, about 40\%.

\begin{figure}
\centering
\includegraphics[width=10cm]{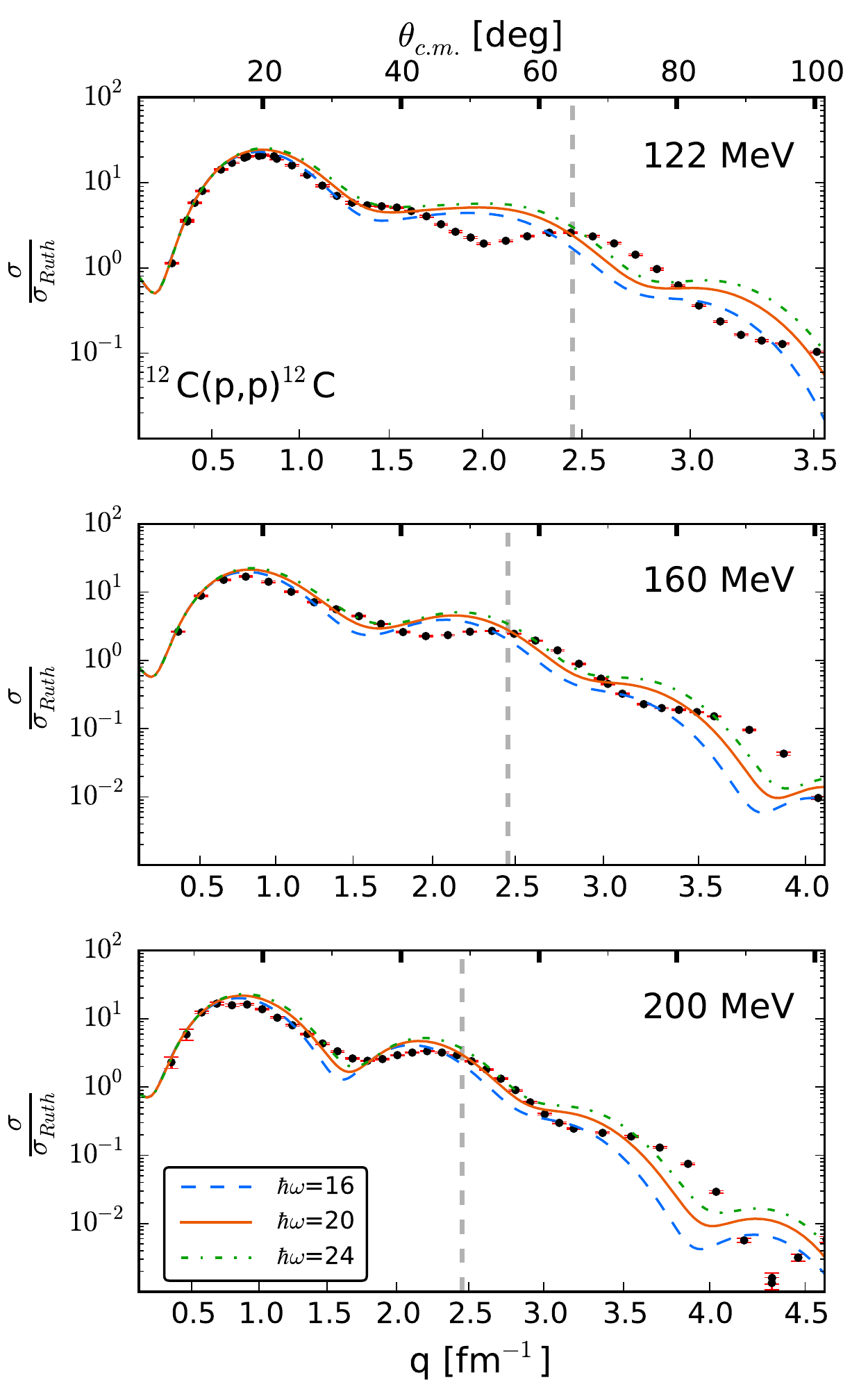}
\caption{The angular distribution of the differential cross section divided by the
Rutherford cross section for elastic proton scattering from $^{12}$C at 122, 160, and
200~MeV laboratory kinetic energy as function of the momentum transfer and the c.m.
angle calculated with the NNLO${_{\rm{opt}}}$ chiral
interaction~\protect\cite{Ekstrom13}. The dashed line represents the calculation based
on nonlocal densities using
$\hbar\omega$~=~16~MeV, the solid line with 20~MeV, and the dash-dotted line with
24~MeV. For all calculations $N_{\rm{max}}$~=~10 is employed. The data for 122~MeV 
are taken from Ref.~\cite{Meyer:1983kd}, for 160~MeV from Ref.~\cite{Meyer:1983kd}, 
and for 200~MeV from Ref.~\cite{Meyer:1981na}. The dashed vertical line in each figure indicates the momentum transfer
$q=2.45$~fm$^{-1}$ corresponding to the laboratory kinetic energy 125~MeV of the $np$
system.
}
\label{fig8}
\end{figure}

\begin{figure}
\centering
\includegraphics[width=10cm]{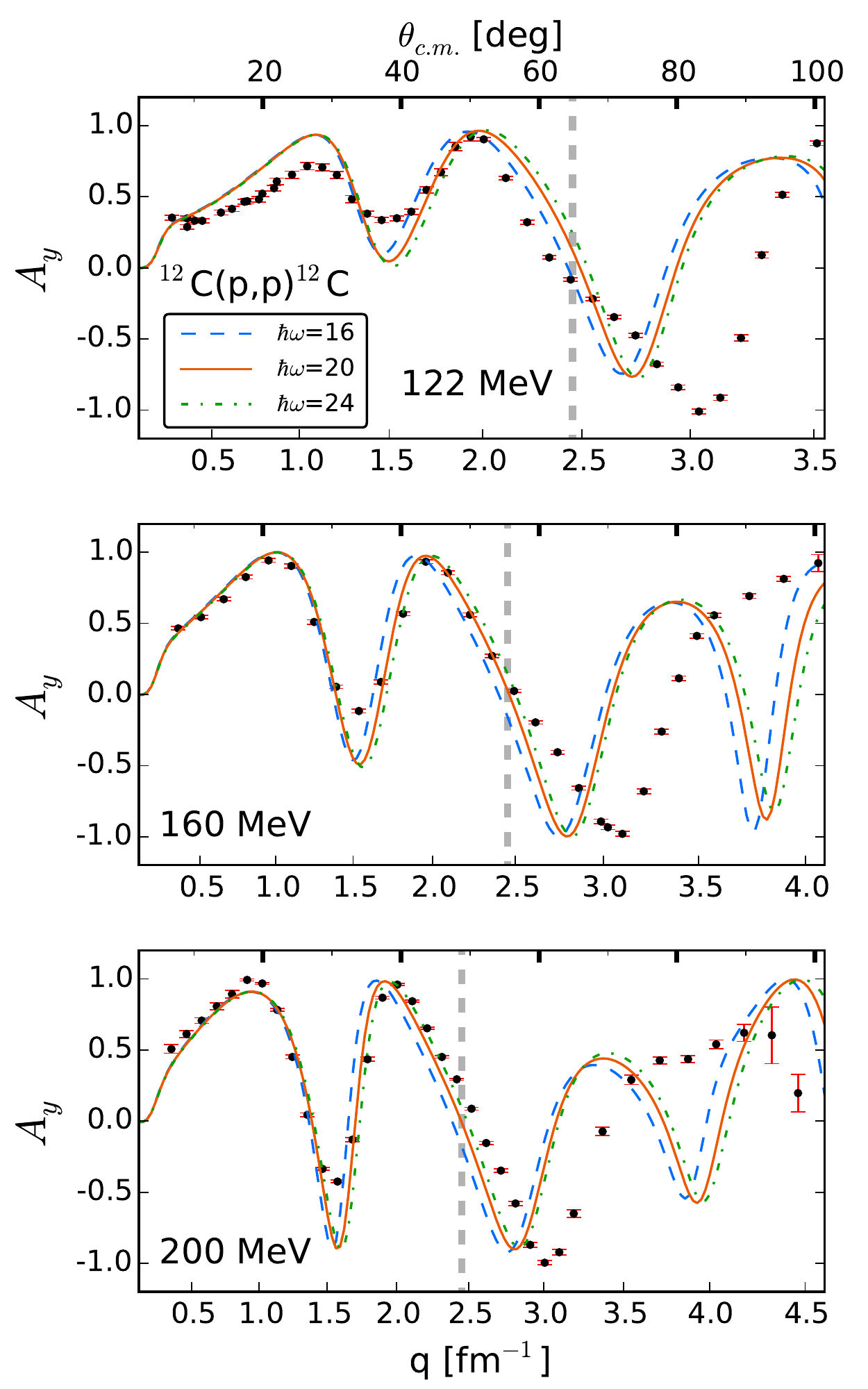}
\caption{The angular distribution of the analyzing power for elastic proton
scattering from $^{12}$C at 122, 160, and
200~MeV laboratory kinetic energy as function of the momentum transfer and the c.m.
angle calculated with the NNLO${_{\rm{opt}}}$ chiral
interaction~\protect\cite{Ekstrom13}. The lines follow the same notation as in
Fig.~\ref{fig9}. The data for 122~MeV are taken from Ref.~\cite{Meyer:1983kd}, for 160~MeV from Ref.~\cite{Meyer:1983kd}, and for 200~MeV from
Ref.~\cite{Meyer:1981na}.
}
\label{fig9}
\end{figure}

The results for proton elastic scattering off $^{12}$C are shown in Figs.~\ref{fig8} and \ref{fig9} for 
laboratory kinetic energies 122, 160, and 200 MeV. The differential cross section divided by the Rutherford
cross section is shown in Fig.~\ref{fig8} while the analyzing power is shown in Fig.~\ref{fig9}. 
Here $N_{\rm max}$ is kept fixed at $N_{\max}=10$ (as for $^{16}$O) while $\hbar \omega$ is varied between 16 and 24 MeV. 
The agreement among the differential cross section experimental data and the calculations is good in the forward direction, and reasonable 
for 160 and 200 MeV even past the 2.45~fm$^{-1}$ marker to about 3.5 fm$^{-1}$ while for 122 MeV, the cross section begins to deviate at the diffraction minima near 2 fm$^{-1}$. The analyzing power calculations in Fig.~\ref{fig9} reasonably agree with the data for proton energies 160 and 200 MeV for $q$ values that are below the corresponding energy to which the NNLO$_{\rm opt}$ 
was fitted, while the results at lower energies 122~MeV deviate more from the data, but retain
the same general shape as for $^{16}$O.
Overall this result for $^{12}$C is unexpectedly good since its ground state, as mentioned above, has a comparatively large non-zero spin contribution. The reason might be that this contribution is fully treated in this formalism, which has captured most of the physics necessary to describe these scattering observables, whereas the effect of the neglected spin-dependent terms appear to be of secondary importance. Indeed,
it is obvious from the 
differential cross section that there are deficiencies in the description, 
since the experimental minima in the cross section differ from the calculation.

\begin{figure}
\centering
\includegraphics[width=10cm]{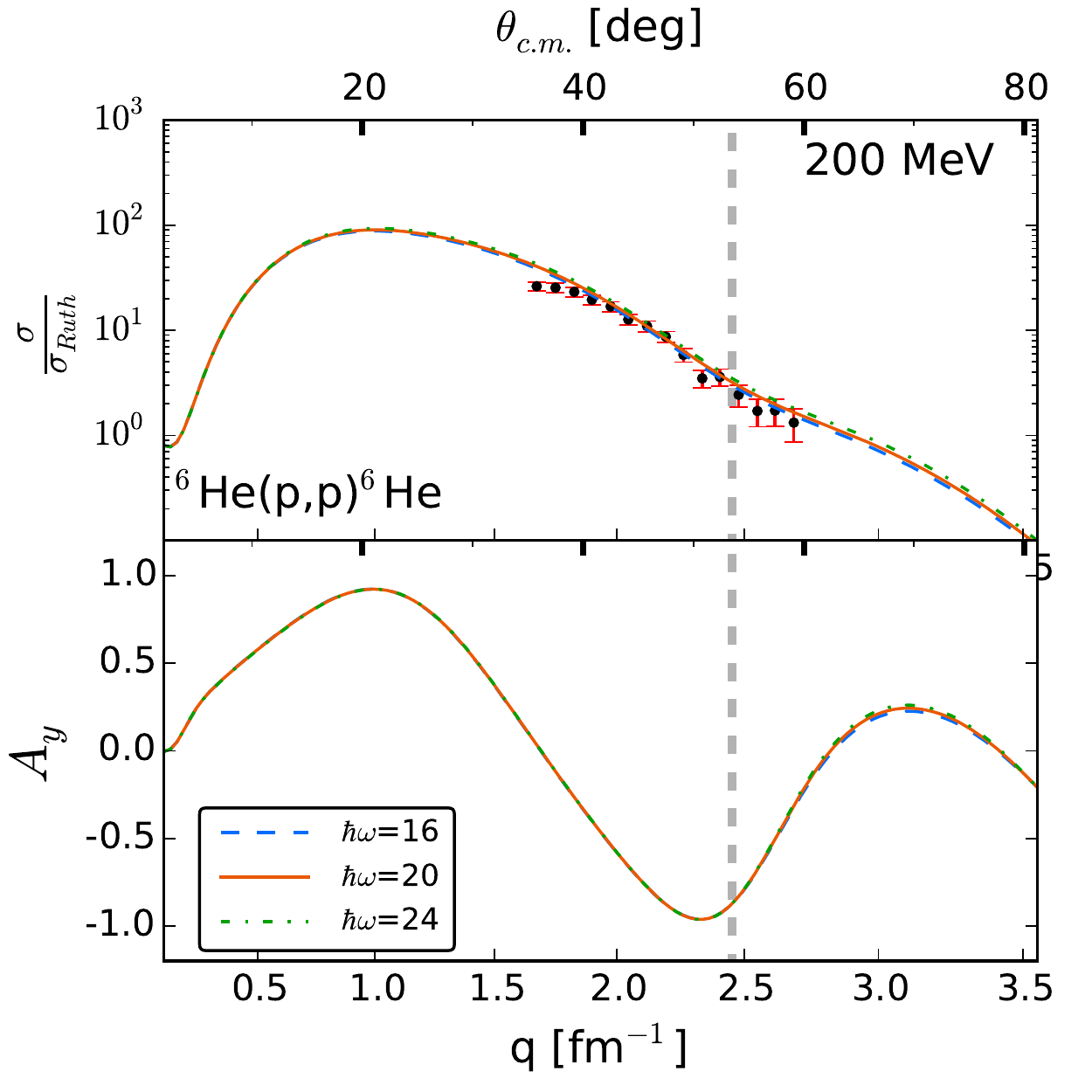}
\caption{The angular distribution of the differential cross section divided by the
Rutherford cross section and the analyzing power for proton scattering from $^6$He at
200~MeV laboratory kinetic energy as
function of the momentum transfer $q$ and the c.m. angle. The dashed line represents
the calculation based
on nonlocal densities using
$\hbar\omega$~=~16~MeV, the solid line with 20~MeV, and the dash-dotted line with
24~MeV. For all calculations $N_{\rm{max}}$~=~18 is employed.
The data are taken from~\cite{Chebotaryov:2018ilv}.
The dashed vertical line in each figure indicates the momentum transfer
$q=2.45$~fm$^{-1}$ corresponding to the laboratory kinetic energy 125~MeV of the $np$
system.
}
\label{fig10}
\end{figure}

Recently the differential cross section of protons off $^6$He has been measured at
200~MeV/nucleon~\cite{Chebotaryov:2018ilv}. Since this energy falls within the range of energies studied here, we
show in Fig.~\ref{fig10} a comparison of the experiment with our calculation of the differential cross
section. Our calculations are performed at $N_{\max}=18$ (same as for $^{4}$He) while $\hbar\omega$ is varied between 16 and 24 MeV, and our results are in good agreement with the available data. In addition we show a prediction of the analyzing power.
Elastic scattering of $^6$He off a polarized proton target
has a somewhat longer history. The first measurement of the analyzing power involving elastic scattering
of an exotic nucleus was carried out at 71~MeV/nucleon~\cite{Sakaguchi:2011rp} and still deviates
considerably from microscopic
calculations~\cite{Kaki:2012hr,Weppner:2011px,Orazbayev:2013dua,Karataglidis:2013zva}. 
Therefore, it will be illuminating to compare our prediction with the
measurement at 200 MeV/nucleon, once fully analyzed~\cite{Sakaguchi:private}.


\subsection{Total and Reaction Cross Sections}

In addition to differential cross sections and spin observables, it is often illuminating to
consider e.g. neutron total cross sections or reaction cross sections since they are integrated over
all scattering angles and may reveal averaged information about the reaction.
In our calculations the total cross section is computed from the imaginary part of the forward
scattering amplitude, while the reaction cross section is obtained using the optical theorem.

\begin{figure}
\centering
\includegraphics[width=1\textwidth]{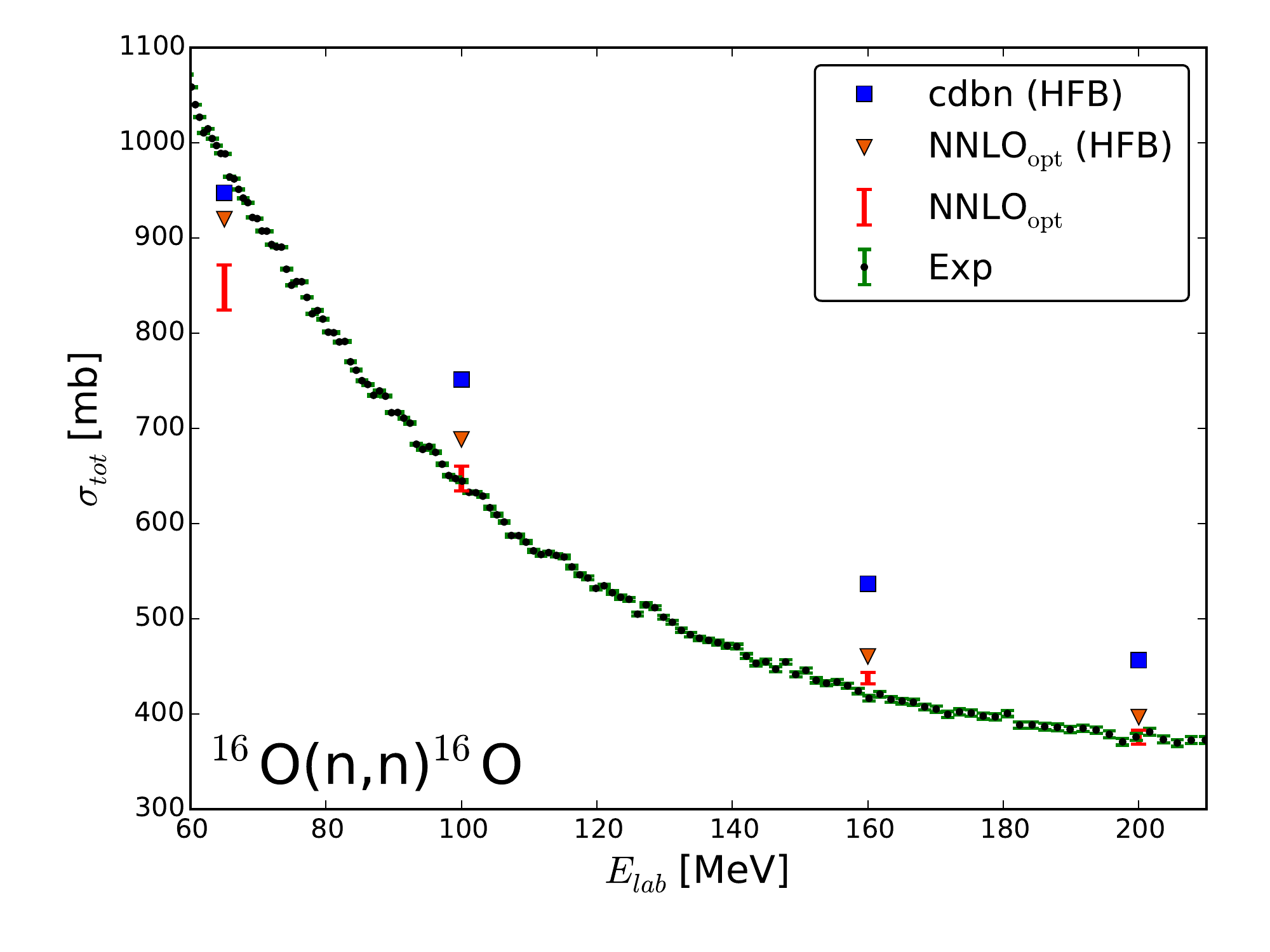}
\caption{The total cross section for neutron scattering from $^{16}$O as function of
the neutron incident energy. The data are taken from Ref.~\cite{Finlay:1993hk}. The
solid band corresponds to calculations using the NNLO${_{\rm{opt}}}$
chiral interaction~\protect\cite{Ekstrom13} consistently in the nonlocal density as
well as in the NN t-matrix\, with the band width determined by
different $\hbar\omega$ values. The downward triangles use the NNLO${_{\rm{opt}}}$ interaction only in the NN t-matrix, while employing a HFB density based on the Gogny-D1S interaction~\cite{Gogny}. The squares use this density together with the CD-Bonn~\cite{Machleidt:2000ge} NN t-matrix.
}
\label{fig12}
\end{figure}

The total cross section for neutron scattering off $^{16}$O is shown in Fig.~\ref{fig12} as function
of the projectile laboratory kinetic energy.
Our calculations between 65 and 200~MeV using values of $\hbar \omega$ between 16 and 24~MeV are shown 
as error bar (without a midpoint). To have a better comparison with previous work using the same
theoretical approach but different input we show as solid squares calculations based on a
Hartree-Fock-Bogoliubov (HFB) nonlocal density with the Gogny-D1S interaction \cite{Gogny} and 
scattering amplitudes from the CD-Bonn potential~\cite{Machleidt:2000ge}. The solid triangles use the
same HFB density but the NNLO$_{\rm opt}$ interaction for the scattering amplitudes. From a comparison of
those three calculations we can conclude that the choice of interaction has a major influence on the value of
the total cross section. However, only the consistent use of the NNLO$_{\rm opt}$ interaction for the
scattering amplitudes and the one-body density leads to a very good agreement with experiment between 100 and
200~MeV. We observe that the calculation at 65~MeV significantly
underestimates the data, indicating that a first-order folding potential is no longer sufficient to describe
the scattering data below about 100~MeV most likely due to a lack of absorption in the single
scattering term. We have found that if one multiplies the effective potential by the scalar $e^{0.244i}$, which
is consistent with similar factors found in Ref.~\cite{Farag:2013ixa}, that it uniformally improves all observables
in which experimental data exists (i.e. reduces the $\chi^2$/datum). We leave an analysis of this effect
 to future work.

\begin{figure}
\centering
\includegraphics[width=1\textwidth]{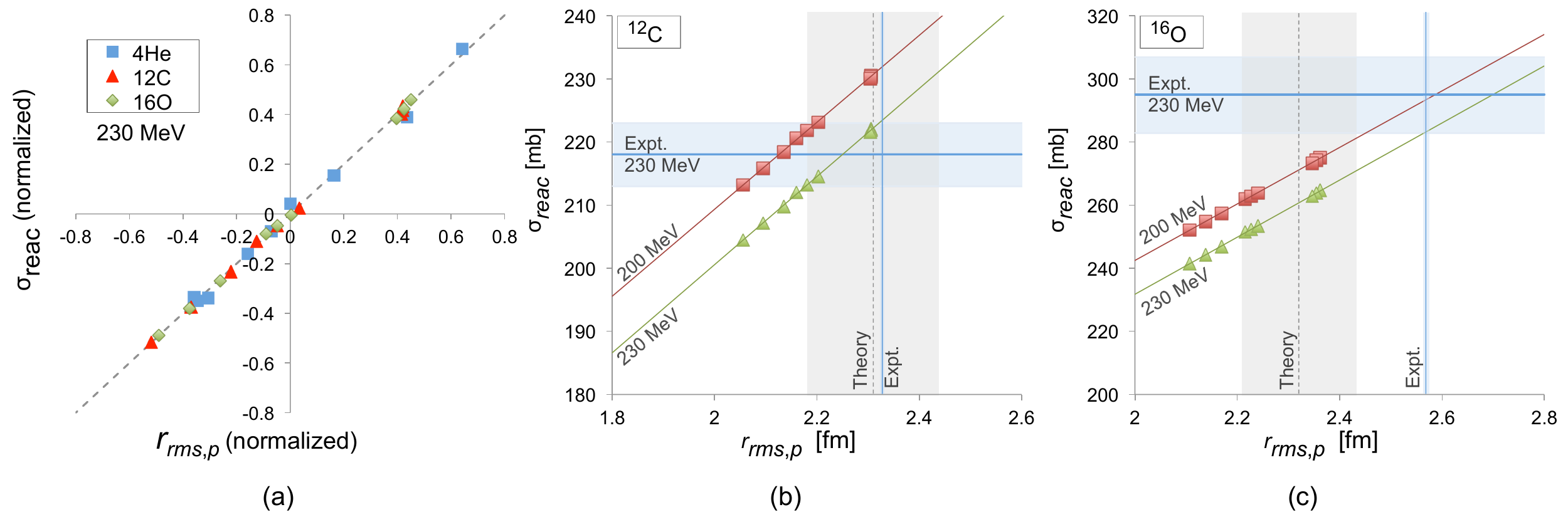}
\caption{Calculated reaction cross sections vs.~ calculated point-proton rms radii $r_{rms}$
for proton scattering at 200~and 230~MeV laboratory projectile kinetic energies off
$^4$He, $^{12}$C, and
$^{16}$O: (a) Correlation plot between the two observables at 230~MeV energy and targets of
$^4$He, $^{12}$C, and
$^{16}$O; to guide the eye, the perfect correlation is indicated by the grey dashed line (see text for details).
(b) and (c) Calculated cross sections as function of point-proton $r_{rms}$ radii for
targets of $^{12}$C (b) and $^{16}$O (c), shown together with point-proton $r_{rms}$
radii extracted from NCSM calculations (labeled as ``Theory"), and compared to
experimental cross sections (where data are available) and experimentally deduced
point-proton rms radii extracted from Ref. \cite{Angeli:2013epw} (labeled as ``Expt."),
with the corresponding errors shown by shaded areas (see text for details). For each
nucleus, calculations are performed for $N_{\rm max}$=6, 8, and 10, and for $\hbar
\omega=$16, 20, and 24 MeV.
}
\label{fig:corr}
\end{figure}

Furthermore, it is worthwhile 
investigating if there is a correlation between observables computed within the structure calculation, and 
cross sections obtained from scattering. Here we use proton scattering data and calculations, since 
neutron total cross section data for $^4$He were not available to us. 
In Table \ref{table1} the total cross section, $\sigma_{tot}$, and the reaction cross section,
$\sigma_{reac}$, for proton scattering at 230~MeV laboratory projectile kinetic energy from $^{16}$O,
$^{12}$C, and $^4$He are given 
 together with the point-proton root-mean-square ($r_{rms,p}$) radii of those nuclei, and compared to experimental data where available . The experimentally deduced point-proton $r_{rms,p}$ are calculated from experimental charge radii \cite{Angeli:2013epw}, using proton and neutron mean-square charge radii $R^2_p = 0.769(13)$ fm$^2$ \cite{Nakamura:2010zzi} and $R^2_n= -0.1149(27)$ fm$^2$ \cite{Angeli:2013epw}, respectively, and a first-order relativistic correction of 0.033 fm$^2$. The proton total cross section refers here to the extracted nuclear part~\cite{Carlson:1996ofz}. Three different values for $\hbar\omega$ are listed in the table, for which $N_{\rm max}$ is kept fixed at values given in the table caption. The calculated total and reaction cross sections are in a close agreement with the data within its error bars, whereas the point-proton rms radii are slightly underpredicted, as is often the case for radii calculated from chiral potentials \cite{PhysRevC.91.051301}.
The table hints at a correlation between the structure and reaction observables.
If one represents the calculated results for each observable as coordinates of a vector,
the scalar product of the two traceless normalized (shifted so the mean of the distribution is zero and the standard deviation is one) vectors is a measure of their correlation \cite{LauneySDD_CPC14}.  Fig. \ref{fig:corr} (a) plots the coordinates of the
traceless normalized vectors corresponding to the reaction cross section ($y$ axis) and to the point-proton rms radius ($x$ axis) for a given nucleus.
Indeed, there is almost perfect correlation between the calculated reaction cross sections with the calculated point-proton rms radii (or equally, the charge radii) for varying NCSM model parameters, $N_{\rm max}$ and $\hbar \omega$, as shown in Fig. \ref{fig:corr} (a). This correlation holds for both ``closed shell" and ``open shell" nuclei under consideration, as well as for different laboratory projectile kinetic energies (only 230~MeV is shown in the figure). This means that the reaction cross section is sensitive to the average radius, and not to the details of the spatial distribution, e.g., the deformation that is pronounced in $^{12}$C. Furthermore, such a feature is especially important for uncertainty quantification of the calculated cross section based on uncertainties obtained for the ground-state rms radius of each nucleus. Calculated cross sections as function of point-proton $r_{rms,p}$ radii for targets of $^{12}$C and $^{16}$O are shown in Figs. \ref{fig:corr} (b) and (c) together with point-proton $r_{rms,p}$ radii extracted from NCSM calculations based on the crossover point as described in Ref.~\cite{Caprio:2014iha}. While not evident from the correlation results,  Figs. \ref{fig:corr} (b) and (c)  reveal a linear dependence  with a comparable slope for  laboratory projectile kinetic energies between 100-230 MeV (as an example, 200~MeV is also shown in the figure). Extracted radii and uncertainties are determined from NCSM calculations up through $N_{\rm max}=10$ and over the $\hbar \omega$ range of 16-24 MeV that contains the fastest rate of convergence of $r_{rms,p}$ with  respect to $N_{\rm max}$, with a rather conservative estimate for the error arising from $\hbar \omega$ variations. Thus, e.g. for $^{12}$C, the extracted ground-state $r_{rms,p}$ of 2.31(13) fm yields an estimated reaction cross section of 222(9) mb for 230 MeV laboratory projectile kinetic energy. It is interesting to note that the extracted $r_{rms,p}$ radius and the estimated reaction cross section lie quite close to the experimental values and agree within the errors. Similarly, for $^{16}$O , for which the extracted ground-state $r_{rms,p}$ is 2.32(11) fm, leading to the estimate for the 230-MeV reaction cross section of 261(10) mb. Such an almost perfect correlation with the rms radii (charge radii) is also observed for the extracted total cross section.

\section{Conclusions and Outlook}
\label{conclusions}

We have calculated the full-folding integral for the first-order effective (optical) potential for NA scattering 
within the framework of the spectator expansion of multiple scattering theory. Those potentials are calculated {\it ab
initio}, i.e. are based consistently on one single NN interaction, in our case the chiral next-to-next-to-leading order
NNLO$_{\rm{opt}}$ interaction from Ref.~\cite{Ekstrom13}, which is fitted to NN data up to 125~MeV laboratory kinetic energy
with $\chi^2 \approx 1$ per degree of freedom, and which describes the 
$A$~=~3, 4 nucleon systems such that the contributions of the 3NFs
are smaller than in most other parameterizations of chiral interactions. Based on this interaction, 
the one-body nonlocal nuclear densities are calculated for the ``closed shell" nuclei $^4$He and $^{16}$O, as well as for the ``open shell" nuclei $^6$He and $^{12}$C using two-body interactions only. The nonlocal densities are created translationally invariant as laid out in Ref.~\cite{Burrows:2017wqn}. Recoil and frame transformation factors are implemented in the calculation of the scattering observables in their complete form. 

We calculated proton elastic scattering observables for the above-mentioned nuclei at laboratory projectile energies from 100 to 200~MeV,  compared them to experimental information, and find them in very good agreement 
with the data in the
angle and momentum transfer regime where the first term of the full-folding effective potential should be valid. Specifically we want to point out the excellent
agreement of the predictions in this regime for the analyzing powers with the data. 
That may be due to the specific fit of the
NNLO$_{\rm{opt}}$ interaction, which seems to slightly change the ratio of the central depth of the effective potential to
its spin-orbit part in addition to minimizing 3NF contribution. The first-order term in the multiple scattering expansion does
not explicitly contain any 3NF contributions, thus the choice of the NNLO$_{\rm{opt}}$ works well with the theoretical
content of the effective potential. Further studies with different interactions
in the future will have to shed more light on the effect including 3NFs in the one-body density for the
first-order effective potential.This will be particularly interesting, since the description of the analyzing powers in the same energy regime is quite different in Ref.\cite{Gennari:2017yez} when the same nuclei are considered.

The theoretical derivation of the first-order potential neglects spin-dependent terms that vanish for nuclear states with total intrinsic spin zero, thus we first considered the ``closed shell'' nuclei $^4$He and $^{16}$O in our study. Since the same formulation is often also applied to ``open shell'' even-even nuclei like $^{12}$C~\cite{Gennari:2017yez,Chinn:1993zza}, we tested our approach also for this case. We find that the description of the differential cross section and the analyzing power is of similar quality as the one we found for $^{16}$O. We also predict differential cross section and analyzing power at 200~MeV for $^6$He, a reaction measured and still being analyzed at RIKEN. Applying a formulation of the first-order term in the multiple scattering theory in which only the NN Wolfenstein amplitudes $A$ and $C$ enter, implies neglecting contribution that come from the other spin couplings inherent in the NN interaction. They may be small, considering that the one-body densities of the nuclei considered are dominated by spin-zero components and also hinted by the reasonably good results presented here for $^{12}$C, but nevertheless this approximation will have to be tested in future work. Exotic nuclei may very well have larger non-zero spin components.

We also calculated total cross sections for neutron scattering and reaction cross sections for proton scattering. We
found that the neutron total cross section for $^{16}$O computed consistently with the NNLO$_{\rm{opt}}$ interaction
gives a superior description of the data compared to previous calculations, which employed 
different interactions for the one-body density and the two-body t-matrix. When comparing total reaction cross
sections with point-proton $r_{rms}$ radii extracted from the structure calculation, we find an almost perfect
correlation between those two quantities for both, ``closed shell''  and ``open shell'' nuclei under consideration,
indicating that the reaction cross section obtained from the first-order folding potential is mainly sensitive to the
average radius of these nuclei.



\appendix

\section{Center-of-Mass (CoM) Contribution in Scattering Observables}
\label{appendixA}

\begin{figure}
\centering
\includegraphics[width=1\textwidth]{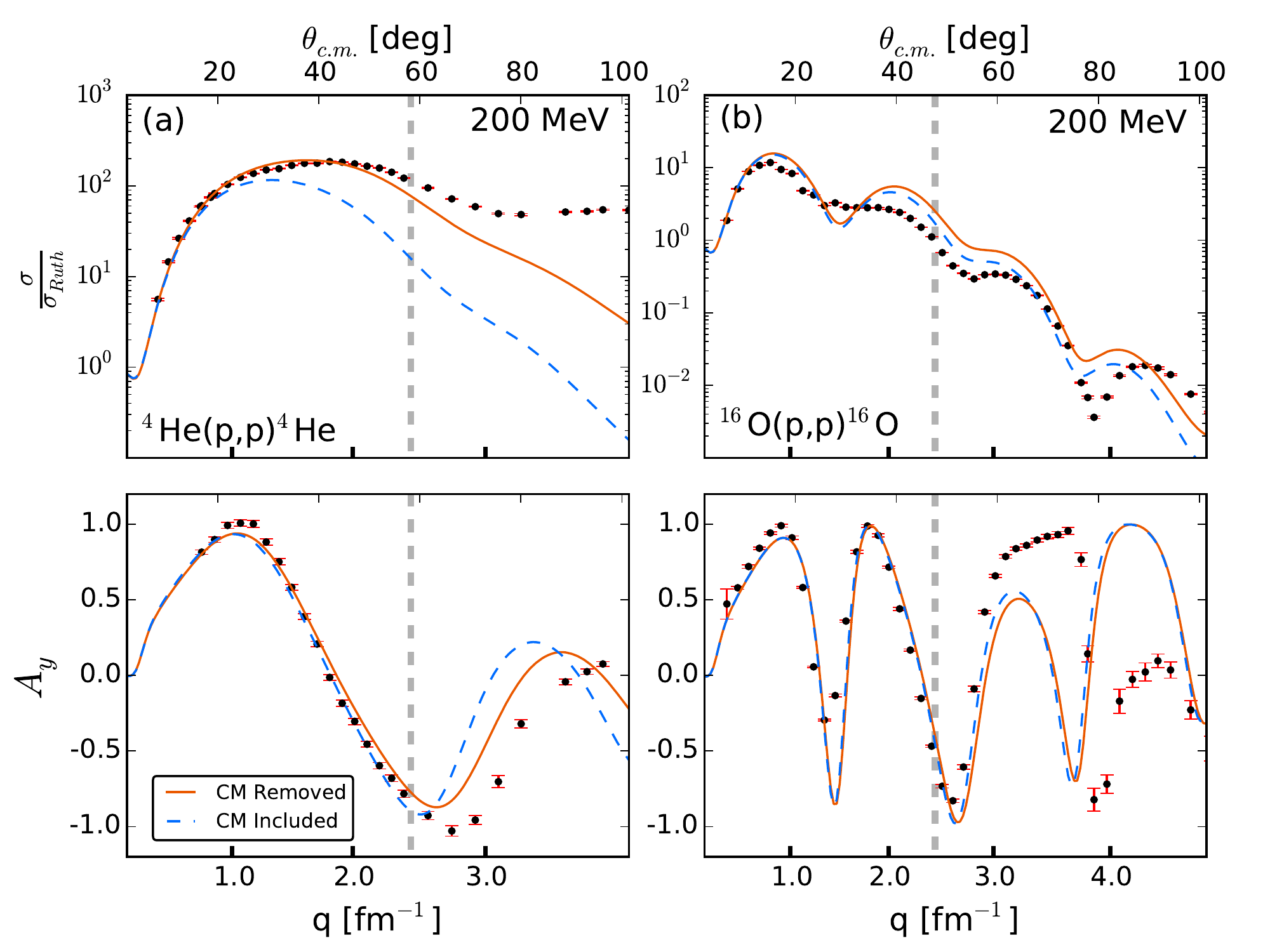}
\caption{The angular distribution of the differential cross section divided by the
Rutherford cross section and the angular distribution of the analyzing power for elastic proton scattering from $^{4}$He (a) and $^{16}$O (b) at 200~MeV laboratory kinetic energy  as function of the momentum transfer and the c.m. angle calculated with the NNLO${_{\rm{opt}}}$ chiral
interaction~\protect\cite{Ekstrom13}. The solid line represents the calculation based
on nonlocal densities without the center-of-mass contribution while the dashed line includes the center-of-mass contribution.
For all $^{4}$He calculations, $N_{\rm{max}}$~=~18 and $\hbar\omega$~=~20~MeV are employed, while all
$^{16}$O calculations employ $N_{\rm{max}}$~=~10 and $\hbar\omega$~=~20~MeV. The data for $^{4}$He at 200~MeV
are taken from~Ref.~\cite{Moss:1979aw} while the data for $^{16}$O at 200~MeV are taken
from~Ref.~\cite{Glover:1985xd}.
The dashed vertical line in each figure indicates the momentum transfer $q=2.45$~fm$^{-1}$ corresponding to
the laboratory kinetic energy 125~MeV of the $np$ system.
}
\label{fig14}
\end{figure}

	It is long-standing knowledge that nuclear one-body densities computed in fixed coordinates, either local or nonlocal, must have their CoM contribution removed in order to be translationally invariant~\cite{DytrychHLDMVLO14,Mihaila:1998qr,Navratil:2004dp,Cockrell:2012vd,Tassie:1958zz,Burrows:2017wqn}.
Working with translationally invariant one-body densities is particularly important in reaction calculations, since those
are carried out in the c.m. frame of the particles involved in the reaction. It is well understood that the size of the CoM contribution decreases with the nuclear mass as $1/A$.  In Fig.~\ref{fig14} the differential cross section divided by the Rutherford cross section along with the analyzing power 
is shown for both $^{4}$He (a) and $^{16}$O (b) at 200 MeV laboratory kinetic energy. The solid lines represent the
full-folding calculation using a translationally invariant nonlocal density, while the dashed lines
represent a calculation containing the CoM contribution. The cross sections follow the expected trend, with $^{4}$He being greatly affected 
already at relatively low momentum transfers, while the effect for $^{16}$O is only evident at large momentum transfers.

The analyzing powers are less affected by the CoM contribution, even for $^{4}$He, which 
is most likely due to the fact that the analyzing powers  are ratios of spin-dependent cross sections,
and deviations in their magnitude are divided out. A similar, even more detailed study is presented in
Ref.~\cite{Gennari:2017yez}. We want to confirm those results and suggest that the
analyzing power should be generally unaffected by the CoM contribution for nuclei $A \gtrsim 16$, while cross sections should
be unaffected for $A \gtrsim 20$. Thus, {\it ab initio} structure calculations for heavier nuclei for which it is not possible to
remove the CoM contribution exactly, can also provide one-body densities for NA scattering calculations.

\begin{acknowledgments}
The authors thank A. Ekstr\"om for sharing the code for the chiral
interaction NNLO$_{\rm{opt}}$ with us.
P. Maris thanks the Funda\c{c}\~{a}o de Amparo \`{a} Pesquisa do Estado de S\~{a}o Paulo (FAPESP) for support under grant No 2017/19371-0.
This work was performed in part under the auspices of the U.~S. Department of Energy under contract
Nos. DE-FG02-93ER40756 and DE-SC0018223, 
by the U.S. NSF (OIA-1738287 \& ACI-1713690),
and of DFG and NSFC through funds provided to the
Sino-German CRC 110 ``Symmetries and the Emergence of Structure in QCD" (NSFC
Grant No.~11621131001, DFG Grant No.~TRR110). 
The numerical computations benefited from computing resources provided
by Blue Waters (supported by the U.S. NSF, OCI-0725070 and ACI-1238993, and the state of Illinois), as well as the Louisiana Optical
Network Initiative and HPC resources provided by LSU ({\tt www.hpc.lsu.edu}), together with resources of the National Energy Research Scientific Computing Center, a DOE Office of Science User Facility supported by the Office of Science of the U.S. Department of Energy under contract No. DE-AC02-05CH11231. 

\end{acknowledgments}

\bibliography{denspot,clusterpot,ncsm}

\clearpage

\begin{table}[!h]
\begin{tabular}{|c|c|c|c|c|c|c|c|c|}
\hline
Target & E [MeV] & Exp. $\sigma_{tot}$[mb] & $\sigma_{tot}$[mb] & Exp. $\sigma_{reac}$[mb] & $\sigma_{reac.}$[mb] & Exp. $r_{rms,p}$ [fm] & $r_{rms,p}$ [fm] & $\hbar \omega$ [MeV] \\
\hline
				&			&								& 359.2 &							& 262.9 &		& 2.346 & 16 \\
$^{16}$O & 230 & 380 $\pm$ 15 	& 351.5 & 295 $\pm$ 12 & 253.3 &	2.569 $\pm$ 0.006	& 2.240 & 20 \\
				&			 &								& 346.2 &							& 246.8 &		& 2.169 & 24 \\
\hline
				&			&							& 288.7 &							& 221.6 &		&  2.304 & 16 \\
$^{12}$C & 230 & 290 $\pm$ 12 & 283.3 & 218 $\pm$ 5 & 214.5 &	2.327 $\pm$ 0.004	& 2.202 & 20 \\
				&			&							& 279.7 &							& 209.8 &		& 2.135 & 24 \\
\hline
				&			&							& 111.4 &	& 86.5 &		&  1.440 & 16 \\
$^{4}$He & 230 & 109 $\pm$ 1 & 111.1 & - & 86.3 & 1.456 $\pm$ 0.005\footnote{A discrepancy between this value and that listed in Ref. \cite{Lu:2013ena} is mainly caused by a difference of the $^{4}$He charge radii 
used here and in Ref. \cite{Lu:2013ena}. However, both numbers agree within error bars.} & 1.437 & 20 \\
				&			&							& 110.8 &	& 86.1 &		& 1.436 & 24 \\
\hline
\end{tabular}
\caption{The total cross section, reaction cross section, and point-proton rms radii for $^{16}$O, $^{12}$C,
and $^{4}$He over a range of oscillator parameter $\hbar \omega$ values. All calculations are performed with
$N_{\rm max}$=10 for $^{16}$O and $^{12}$C while $^{4}$He used $N_{\rm max}$=18. The experimental total cross
section and reaction cross section values are taken from \cite{Carlson:1996ofz}. The total cross section is
an extracted value for the nuclear part. The experimentally deduced point-proton rms radii are extracted from
\cite{Angeli:2013epw}.
} 
\label{table1}
\end{table}

\clearpage


\end{document}